\documentclass[reprint,amsmath,amssymb,aps,pra,nofootinbib]{revtex4-2}

\usepackage{graphicx}
\graphicspath{{./figs/}}
\usepackage{dcolumn}
\usepackage{bm}
\usepackage[breaklinks=true,colorlinks=true,
linkcolor=blue,urlcolor=blue,citecolor=blue]{hyperref}
\usepackage{longtable}
\usepackage{multirow}
\usepackage[dvipsnames]{xcolor}
\usepackage[pscoord]{eso-pic}
\usepackage[normalem]{ulem}

\newcommand{\ie}{i.e.,~}
\newcommand{\eg}{e.g.,~}

\newcommand{\msun}{\text{M}_\odot}

\newcommand{\rchitwo}{\tilde{\chi}^2}

\newcommand{\placetextbox}[3]{
	\setbox0=\hbox{#3}
	\AddToShipoutPictureFG*{
		\put(\LenToUnit{#1\paperwidth},\LenToUnit{#2\paperheight})
                {\vtop{{\null}\makebox[0pt][c]{#3}}}
	}
}

\begin{document}

\placetextbox{0.90}{0.985}{\normalsize \normalfont ULB-TH/20-04}
\placetextbox{0.915}{0.97}{\normalsize \normalfont TTK-20-09}

\title{Shedding light on the angular momentum evolution of binary neutron
star merger remnants: a semi-analytic model}

\author{
	Matteo Lucca,$^{1}$
	Laura Sagunski,$^{2,3}$
	Federico Guercilena,$^{4}$ and
	Christian M. Fromm$^{2,5}$
	\\
	\vspace{0.15 cm}
	$^{1}$\textit{\small Service de Physique Th\'{e}orique,
        Universit\'{e} Libre de Bruxelles, C.P. 225, B-1050 Brussels,
        Belgium}\\
	$^{2}$\textit{\small Institute for Theoretical Physics,
		Goethe University, Frankfurt am Main, 60438, Germany}\\
	$^{3}$\textit{\small Institute for Theoretical Particle Physics
        and Cosmology, RWTH Aachen University, Aachen, 52074, Germany}\\
	$^{4}$\textit{\small Technische Universit\"at Darmstadt,
		Institut f\"ur Kernphysik, Schlossgartenstr. 2, 64289 Darmstadt,
		Germany} \\
	$^{5}$\textit{\small Max-Planck-Institut f\"ur Radioastronomie, 53121,
        Bonn, Germany}
}

\begin{abstract}
    The main features of the gravitational dynamics of binary neutron
    star systems are now well established. While the inspiral can be
    precisely described in the post-Newtonian approximation, fully
    relativistic magneto-hydrodynamical simulations are required to model
    the evolution of the merger and post-merger phase. However, the
    interpretation of the numerical results can often be non-trivial, so
    that toy models become a very powerful tool. Not only do they
    simplify the interpretation of the post-merger dynamics, but also
    allow to gain insights into the physics behind it. In this work, we
    construct a simple toy model that is capable of reproducing the whole
    angular momentum evolution of the post-merger remnant, from the
    merger to the collapse. We validate the model against several fully
    general-relativistic numerical simulations employing a genetic
    algorithm, and against additional constraints derived from the
    spectral properties of the gravitational radiation.
    As a result, from the remarkably close overlap between the
    model predictions and the reference simulations within the first
    milliseconds after the merger, we are able to systematically shed
    light on the currently open debate regarding the source of the
    low-frequency peaks of the gravitational wave power spectral density.
    Additionally, we also present two original relations connecting the
    angular momentum of the post-merger remnant at merger and collapse to
    initial properties of the system.
\end{abstract}

\maketitle

\section{Introduction}
Accurately modeling the structure of neutron stars (NSs) and their dynamics in
binary systems is known to be one of richest problems in physics. This is
because these environments offer the unique possibility to precisely
study the interplay between nuclear physics and general relativity at extreme
densities. In fact, already during the '30s great theoretical effort
was dedicated to NSs and to their nuclear properties, while a renewed interest
followed the observation of the first X-ray sources \cite{PhysRevLett.9.439}
and radio pulsars \cite{1968Natur.217..709H} in the '60s (see e.g.,
\cite{Yakovlev_2013, schofield1982neutron} for dedicated historical overviews).

The role of NS binary systems became relevant after the observation of
the first binary pulsars in 1975 \cite{1975ApJ...195L..51H} and the later
hypothesis that such systems could be ideal sources of short gamma-ray
bursts (SGRBs) \cite{Eichler89, Narayan92}. Following these pioneering
ideas, decades of experimental developments (see e.g.,
\cite{LIGO_history} and \cite{Fermi_GMB_history} for a summarized history
of the LIGO collaboration and the \textit{Fermi}-GBM mission,
respectively) led on August the 17th 2017 to the first multi-messenger
observation of a NS-NS merger event by a worldwide collaboration of
observatories \cite{abbott2017b_NS, abbott2017d_NS}. The remarkable
interplay between the detection of the gravitational wave (GW) signal
GW170817  \cite{abbott2017a_NS}, the slightly delayed SGRB (GRB170817A)
\cite{Goldstein2017, Savchenko2017} and the following UV-optical-NIR
counterpart \cite{Coulter2017, abbott2017c_NS} has proven to be a strong
confirmation of the theoretical framework surrounding NS
binaries.

At the same time, however, when investigating complex systems such as NS
binaries, one faces several challenges like, for instance, solving highly
non-linear relativistic hydrodynamics equations, predicting the spectral \linebreak
evolution of the GW emission and developing adequate numerical~methods.~Building on the first pioneering numerical studies that tried to approach such
difficulties\linebreak with Newtonian or post-Newtonian approximations \cite{Rasio92,
Ruffert1994}, already more than a decade ago several groups star- ted to develop
fully general-relativistic codes \cite{Shibata:1999wm, Shibata01a,
Shibata:2002jb, PhysRevD.65.084024, Shibata:2006nm, Anderson:2007kz,
Yamamoto:2008js, Baiotti08, Rezzolla:2010} in order to precisely explore the
system's complexities. Additional effort has then been dedicated in the
following years to also include the contribution from e.g., magnetic\linebreak fields
\cite{Anderson2008, Giacomazzo2011b, Ciolfi2017}, thermal effects
\cite{Kaplan2013, Endrizzi:2019trv}, initially spinning\linebreak NSs \cite{Kastaun2014},
neutrino cooling \cite{Bernuzzi2015b}, viscosity \cite{Shibata2017b} and
eccentricity \cite{Chaurasia2018}.

Although in some regards these simulations allow for a multitude of very
accurate predictions, a deeper understanding of the connection between
NSs in binary systems and their nuclear properties, i.e., their equation
of state (EOS), has only been achieved in recent years with the help of
several systematic studies (see e.g., \cite{Stergioulas2011b,
Bauswein2012a, Bauswein2012, Hotokezaka2013c, Bauswein2014, Takami2015,
Rezzolla2016}). Of particular interest has been the relation of EOS
dependent quantities, such as the shape of the post-merger remnant, to
spectral observables, such as the peaks in the power spectral density
(PSD).

To summarize some of the main results obtained in these studies (see
e.g., \cite{Rezzolla2016} for a more detailed discussion), it has been
possible to draw the picture of a post-merger evolution divided in two
phases: a so-called transient phase immediately following the merger and
lasting only for few milliseconds, and a so-called quasi-stationary phase
where the hyper massive NS (HMNS) develops a bar-like shape and which
lasts until the collapse occurs.

In most cases, several distinct peaks are clearly recognizable in the PSD.
\cite{Stergioulas2011b} was the first to suggest that the usually largest peak,
commonly labeled $f_2$ in the literature, is related to the quadrupolar $m=2$
mode of the HMNS, characteristic of the quasi-stationary phase. This is also
supported by the fact that this peak is mostly produced after the first 3 ms
after the merger (see e.g., Fig.  9 of \cite{Takami2015}). As an example of the
interplay between spectral properties and the NS EOS, it has also been shown
that this quantity is universally related, i.e., independently of the EOS, to
the compactness \cite{Takami2015} and the tidal deformability
\cite{Bernuzzi:2015rla, Rezzolla2016} of the progenitor~NSs.

However, in particular in the low-frequency region of the spectrum, the
shape of the PSD becomes rather irregular. The physical interpretation of
the origin of these secondary peaks is subject of an ongoing debate. On
one side, it has been suggested that two different effects happening at
the same time could be the source of the low-frequency peaks: the
superposition of radial and quadrupolar pattern \cite{Stergioulas2011b,
Bauswein:2015yca}, and the rotating pattern of a deformation of
the spiral shape \cite{Bauswein:2015yca}. The first one dominates
for high-mass binaries, while the second effect is stronger for low-mass
binaries. On the other side, \cite{Takami2015} proposed an alternative
interpretation where the most prominent low-frequency peak, in this
context labeled $f_1$, is the result of the oscillations between the
stellar cores in the first rotation periods after the merger. This
mechanism would also explain the presence of a third prominent peak at
frequencies higher that $f_2$, usually dubbed $f_3$, and the fact that
$f_2\approx(f_1+f_3)/2$ \cite{Takami2015, Rezzolla2016}.

Since HMNSs are supported against collapse by differential rotation, the
loss of angular momentum in GWs becomes an ideal quantity to investigate
the dynamical properties of the post-merger object such as, for instance,
the evolution of the transient phase. Following this idea, in this work
we aim to construct a simple, but complete toy model which is able to
reproduce the full time evolution of the HMNS angular momentum, from
merger to collapse. As it turns out, this method does not only allow to
shed some light on the controversy regarding the transient phase and the
origin of the secondary peaks, but it also points towards several
original conclusions.

This paper is organized as follows. Sec.~\ref{sec:sample} summarizes the
properties of the data sample to which our model is fitted.
Sec.~\ref{sec:j_evol} describes the mathematical model employed in this
work to reproduce the time evolution of the HMNS angular momentum. In
Sec.~\ref{sec:validity} (and App. \ref{Sec: Appendix A}), we present
the genetic algorithm (GA) which is used to determine the set of free
parameters of the model.  There, we also investigate the validity of the
GA results as well as the predictive power of the model when the GA
results are employed. Finally, we conclude in Sec.~\ref{sec:conclusions}
with a summary and a discussion on the possible applications of our
results.

\textit{Remark on notation:} Throughout this paper, all quantities are
expressed in a geometrized system of units in which $c=G=1$, and we use
the solar mass~M$_{\odot}$ as unit of mass, unless stated otherwise.
Furthermore, the subscript $_{\rm TOV}$ indicates a quantity referring to
the maximum mass (non-rotating) neutron star model for a given EOS, i.e.,
the Tolman-Oppenheimer-Volkoff (TOV) limit, while the subscript $_{\rm
NS}$ is used with respect to the characteristics of the initial NSs in a
binary system.

\section{Data sample}\label{sec:sample}
As data sample to calibrate the free parameters of the model, we use the
numerical results of the direct, large-scale, fully relativistic
simulations of binary neutron star (BNS) mergers performed by
\cite{Takami2015, Rezzolla2016}.

These simulations use a fourth-order finite-difference code, which solves
the Baumgarte-Shapiro-Shibata-Nakamura (BSSN) formulation
\cite{Nakamura87,Shibata95,Baumgarte99,Brown09} of the Einstein
equations. The general-relativistic hydrodynamics equations are solved
using a finite-volume method, employing the Harten-Lax-van Leer-Einfeldt
(HLLE) Riemann solver \cite{Harten83} and the Piecewise Parabolic Method
(PPM) \cite{Colella84} for the reconstruction of the evolved variables.
For the time integration a Method of Lines (MoL) algorithm is used with
the explicit fourth-order Runge-Kutta RK4 method. The interested reader
can find extensive details about the mathematical and numerical setup of
the simulations in \cite{Takami2015, Rezzolla2016} and the references
therein.

While the simulations do properly account for the evolution of spacetime
and the fluid dynamics of the NS matter, several physical effects which
are expected to have a significant impact on the mechanism and time of
collapse of the remnant object are neglected. These include fluid
viscosity, electromagnetic fields and neutrino transport. Also, thermal
effects in the fluid are accounted for only approximately via an
ideal-gas EOS contribution.

Overall, this set of simulations encompasses various BNS models, employing five
different so-called "cold" EOSs, \ie at zero temperature and in
beta-equilibrium, but coupled to an ideal-gas EOS to approximately take into
account thermal effects. The employed EOSs are: ALF2 \cite{Alford2005}, APR4
\cite{Akmal1998}, GNH3 \cite{Glendenning1985}, H4 \cite{Glendenning1992}, and
SLy~\cite{Douchin2001}. As pointed out in \cite{Takami2015}, all of
these EOSs comply with the observational bound on the maximum mass in a NS
found in \cite{Antoniadis1233232} as well as with the more recent results
presented in \cite{Cromartie:2019kug, Linares_2018} at 95\% confidence level (see e.g.,
\cite{Godzieba:2020tjn} for additional discussions on these maximum mass
bounds). For each EOS 9 initial NSs gravitational masses are
considered\footnote{Note that, compared to the sample presented in
\cite{Takami2015,Rezzolla2016}, the data regarding the case of models
\texttt{SLy-M1225}, \texttt{SLy-M1375} and \texttt{SLy-M1400} has been lost
because of a data server breakdown and is therefore not used in the following
calculations.} ($M_{\rm NS}=1.200,$ $1.225,$ $1.250,$ $1.275,$ $1.300,$
$1.325,$ $1.350,$ $1.375,$ $1.400$ $\msun$), thereby covering a wide range in
the NSs compactness.

All BNS models considered are equal mass systems. Indeed, this represents
one of the current limitations to the applicability of our model,
although one that could be easily removed given access to unequal-mass
BNS simulations data. We hope to be able to undertake this effort in the
future.
\begin{figure}[t]
	\centering
	\includegraphics[width=\columnwidth]{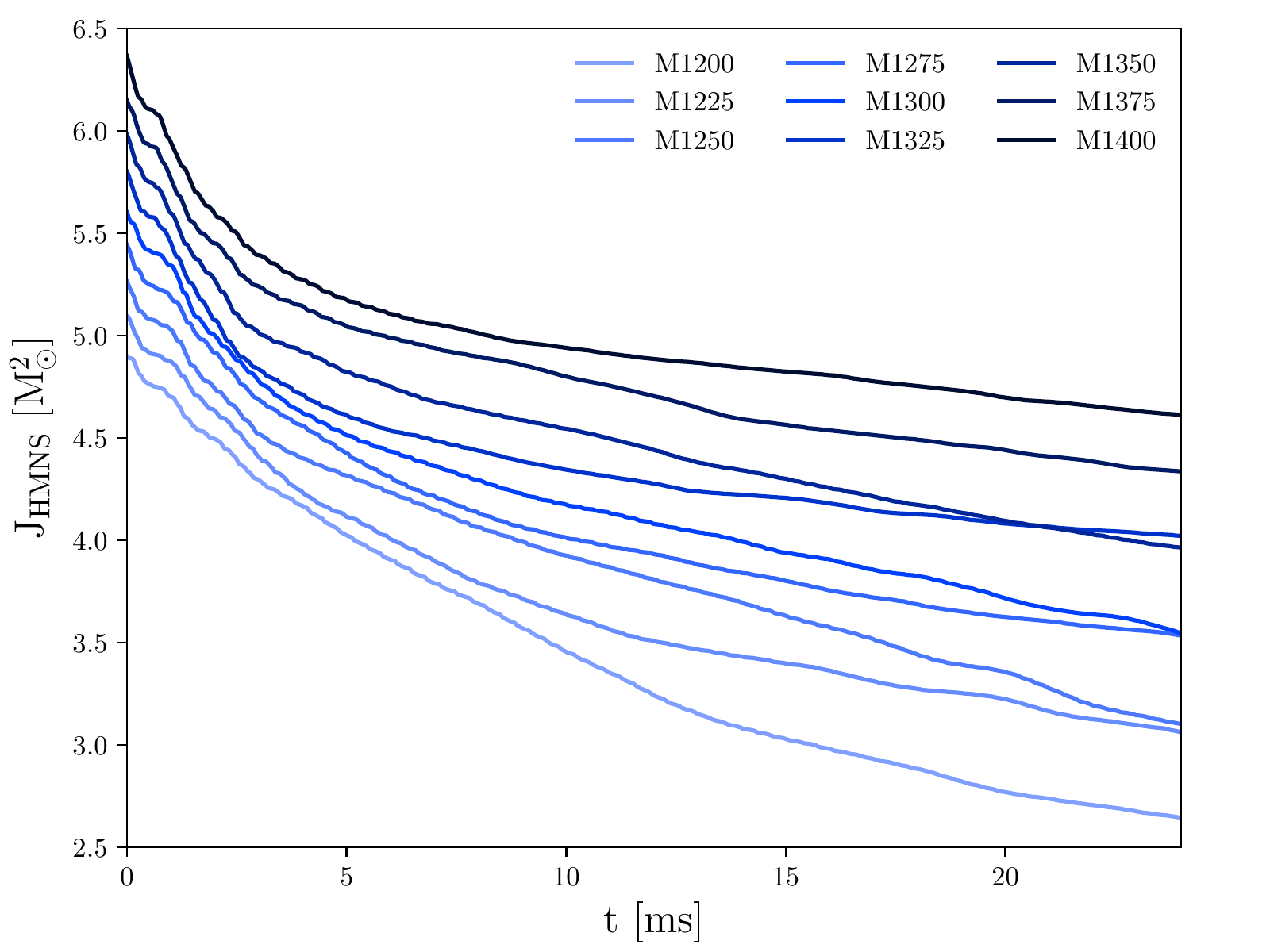}
	\caption{Angular momentum of the merger remnant as a function of
	time in the post-merger phase from the simulations of
	\cite{Takami2015, Rezzolla2016} for the case of the APR4 EOS.
	The origin of the time axis is the instant of merger. Different
	curves correspond to different initial gravitational masses of
	the NSs.}
	\label{fig:j_example}
\end{figure}

\section{Definition of the toy model}
\label{sec:j_evol}
The main quantity that we are interested in modeling from the
aforementioned simulations is the evolution of the angular momentum of
the merger remnant, denoted as $J_{\rm HMNS}$. This is computed from the
simulation data as the difference between the Arnowitt-Deser-Misner (ADM)
angular momentum of the system, $J_{\rm ADM}$ (\ie the initial value of
the angular momentum), and the angular momentum emitted in GW, $J_{\rm
GW}$, \ie ${J_{\rm HMNS}=J_{\rm ADM}-J_{\rm GW}}$ (see \eg
\cite{Baumgarte2010a} for a definition of these quantities). Since we are
only interested in the post-merger behavior, we define the time of merger
$t_{\rm merger}$ as the time at which the GW strain amplitude reaches its
first maximum \cite{Rezzolla2016} and set it from now forth as the origin
of our time axis, \ie $t_{\rm merger}=0$.
\begin{figure}[t]
	\centering
	\vspace{0.3 cm}
	\includegraphics[width=\columnwidth]{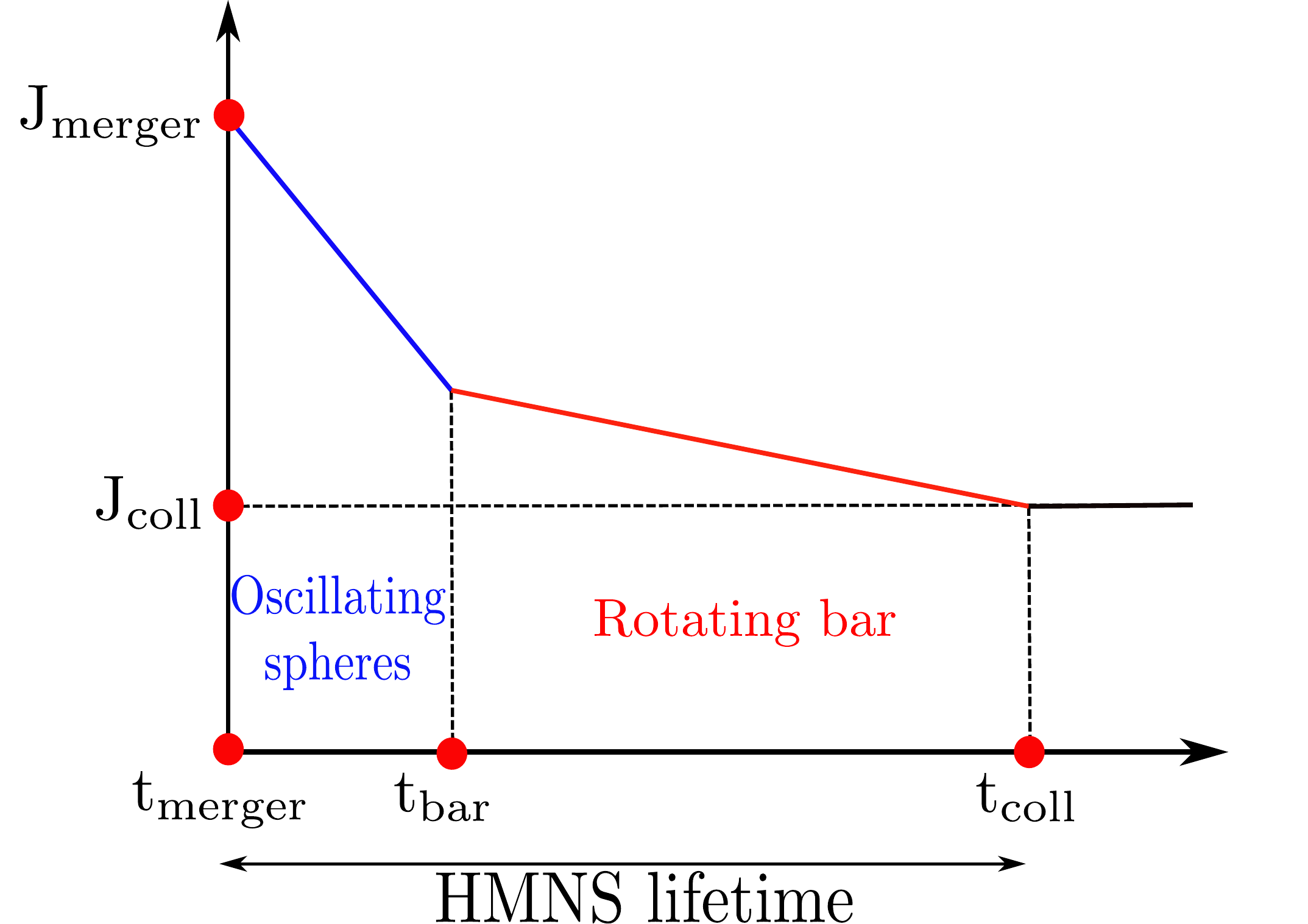}
	\caption{Schematic representation of the post-merger angular
	momentum evolution as a function of time as predicted by the
	model defined in section Sec.~\ref{sec:j_evol}.}
	\label{fig:model}
\end{figure}

As an example of the post-merger angular momentum behavior,
Fig.~\ref{fig:j_example} shows $J_{\rm HMNS}$ as a function of time, as
computed from the simulations of \cite{Takami2015, Rezzolla2016},
employing the APR4 EOS. Note that all the curves can be roughly described
by two lines of constant slope, the first transitioning into the second
at approximately 5 ms after merger. This behavior is schematically shown
in the simple graphical depiction of Fig.~\ref{fig:model}, which
also illustrates the main quantities discussed in the following
paragraphs. In particular, one can follow the angular momentum evolution
starting from the merger, undergoing the transient phase where the
stellar cores oscillates, and finally reaching the rotating bar
configuration, which lasts until the angular momentum is sufficiently
large to compensate the gravitational pressure. A detailed definition of
all the quantities involved in these phases is the goal of this section.

\begin{figure}[t]
	\centering
	\includegraphics[width=\columnwidth]{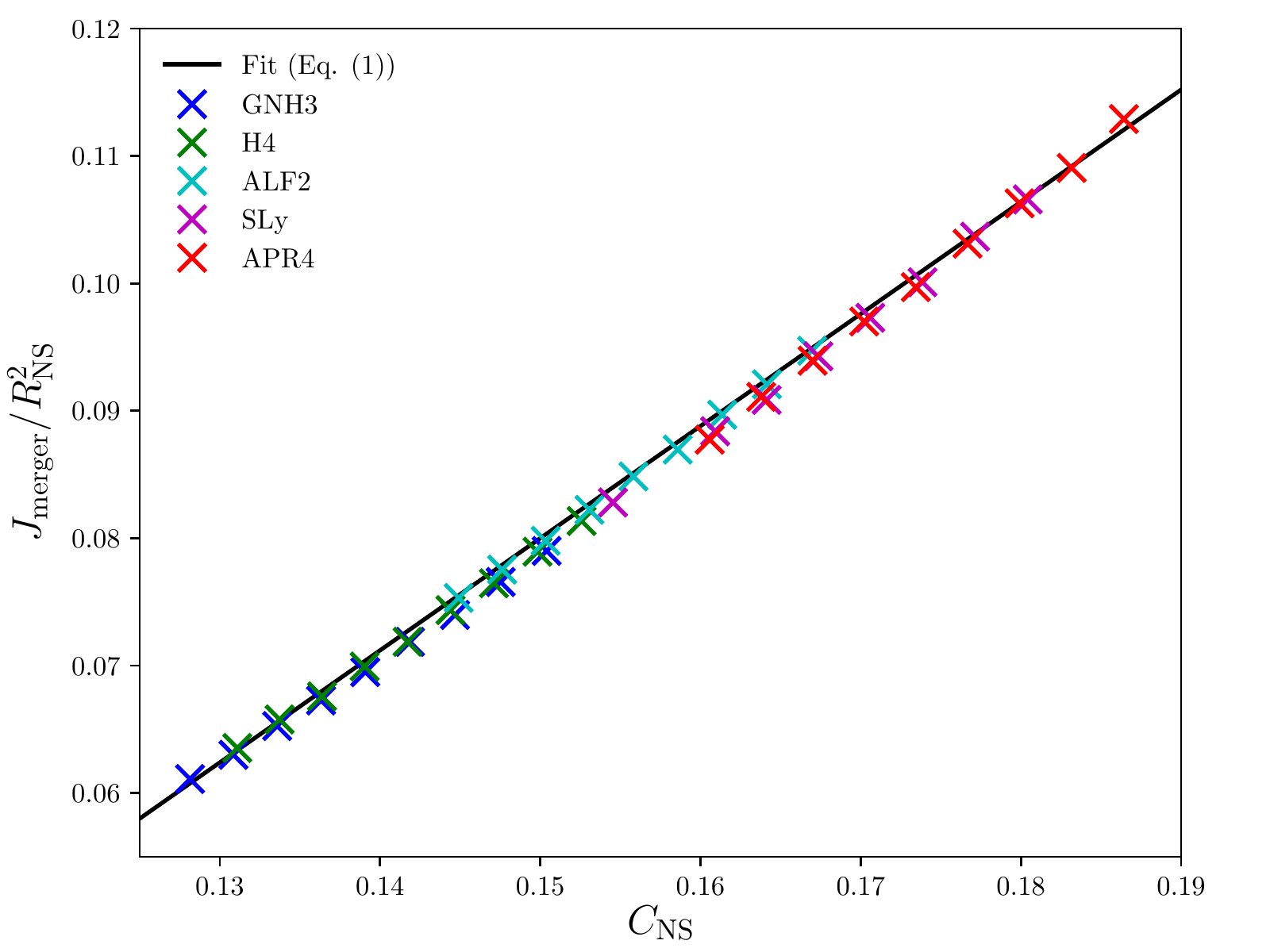}
	\caption{Angular momentum at the time of merger normalized by the
	square of the progenitor NS radius as a function of the
	compactness of the initial NSs.}
	\label{fig:JR_over_C}
\end{figure}

First of all, it is necessary to define the HMNS angular momentum at
merger, i.e., $J_{\rm merger}=J_{\rm HMNS}(t=0)$. We find that the
dimensionless quantity $J_{\rm merger}/R^{2}_{\rm NS}$, where $R_{\rm
NS}$ is the progenitor NS radius, can be fit by a simple linear universal
(i.e., EOS independent) relation in terms of the compactness of the NSs,
$C_{\rm NS}$, namely
\begin{align}\label{eq:Jmerger}
	\frac{J_{\rm merger}}{R^{2}_{\rm NS}}=a_1 C_{\rm NS}+a_2\,,
\end{align}
where the coefficients have values $a_1=0.8765\pm0.0051$ and
${a_2=-(5.209\pm0.077)\times10^{-2}}$. In Eq. \eqref{eq:Jmerger} and
following the errors correspond to $1\sigma$ uncertainties. This
universal relation is shown in Fig.~\ref{fig:JR_over_C} along with the
simulation data.

After having determined the initial value of the angular momentum
evolution in the post merger phase in Eq.~\eqref{eq:Jmerger}, we need to
model the subsequent evolution of $J_{\rm HMNS}$. As mentioned above, to
this end we employ the model first proposed in \cite{Takami2015}, which
we summarize in the following paragraphs, and extend it to later times.

\begin{figure}
	\centering
	\includegraphics[width=0.84 \columnwidth]{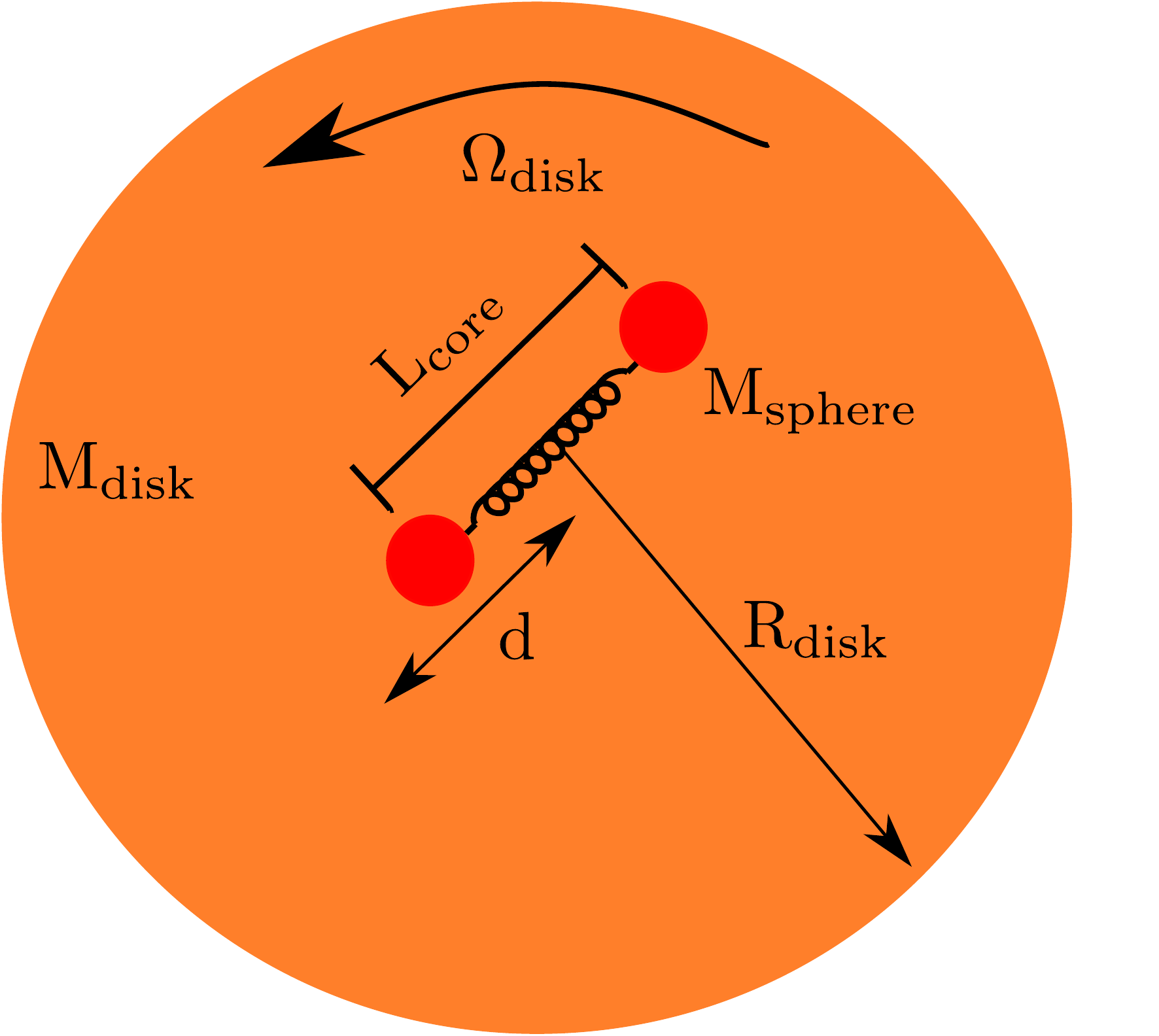}
	\caption{Schematic drawing of the model system for the first
	phase of the post-merger evolution of the merger remnant.}
	\label{fig:spheredisk}
\end{figure}

In the toy model, the HMNS is modeled, in a first phase immediately after
merger, by a very simple mechanical system consisting of two spheres
placed over a rotating disk and connected by a spring (compare to
Fig.~\ref{fig:spheredisk}.). After approximately 5 ms, a second phase
begins where the HMNS is modeled as a rigid rotating bar.

We focus at first on the transient phase. Consider a mechanical system
composed of a disk of mass $M_{\rm disk}$ and radius $R_{\rm disk}$
rotating at a given angular velocity $\Omega_{\rm disk}$ with
corresponding rotation frequency $f_{\rm disk}$. Two spheres are placed
on the disk, each of mass $M_{\rm sphere}$, connected to each other by a
spring and free to oscillate along their separation vector.

Because of angular momentum conservation the angular velocity of the
system increases when the two spheres are closer, and slows down when
they move apart. The rotation frequency of the system is therefore bound
between a minimum value, $f_{\rm disk,1}$, and a maximum one, $f_{\rm
disk,3}$.  Dissipative processes (\ie the friction of the spring between
the sphere in this approximation) damp the oscillations in the spheres'
separation on a timescale of a few milliseconds towards a constant
rotation frequency value $f_{\rm disk,2}\simeq(f_{\rm disk,1}+f_{\rm
disk,3})/2$, which remains nearly unchanged until the collapse time. We
call the sum of the spheres' masses $M_{\rm core}=2M_{\rm sphere}$, and
the separation between the two $L_{\rm core}$. The spring has an elastic
constant $k$ such that $\Omega_{\rm disk}\approx\sqrt{k/M_{\rm core}}$.

With the definitions above, the equation describing the radial
displacement $d(t)$ of a sphere with respect to its position at rest
takes the form \cite{Takami2015}
\begin{align}
	\ddot{d} +\frac{4kd}{M_{\rm core}}
	-\Omega_{\rm disk}^{2}d  +\frac{2b(\dot{d}-v_0)}{M_{\rm core}}=0\,,
	\label{eq:d}
\end{align}
where the angular velocity $\Omega_{\rm disk}$ is a time-dependent
quantity computed as
\begin{align}
	\Omega_{\rm disk}=\frac{J_{\rm merger}/M_{\rm core}}
	{d^2+M_{\rm disk}R^{2}_{\rm disk}/(2M_{\rm core})}\,,
	\label{eq:omega}
\end{align}
$b$ is a damping constant accounting for the dissipative processes
mentioned above, $v_0$ is the speed at plunge (\ie the speed at which the
oscillations of the two cores begin), and overdots indicate
differentiation with respect to time.

Assuming without loss of generality the rotation of the disk to take
place in the $(x,y)$ plane, the non-zero components of the
quadrupole-moment tensor of the model system take the form
\begin{subequations}
	\begin{align}
		& Q^{\rm disk}_{xx}=\frac{M_{\rm core}d^{2}}{2}[1+
		\cos(2\Omega_{\rm disk} t)] \\
		\label{eq:quad.toy1}
		& Q^{\rm disk}_{yy}=\frac{M_{\rm core}d^{2}}{2}[1-
		\cos(2\Omega_{\rm disk}t)] \\
		& Q^{\rm disk}_{xy}=\frac{M_{\rm core}d^{2}}{2}\sin(2\Omega_{\rm disk}t)
		=Q^{\rm disk}_{yx}\,.
		\label{eq:quad.toy2}
	\end{align}
\end{subequations}

From the quadrupole formula, the angular momentum
lost by the system per unit time (in the $z$-direction, given our choice
of orientation of the system) is \cite{Maggiore2007}
\begin{align}
	\frac{dJ}{dt}= \frac{2}{5}\epsilon^{zkl}
	\langle\ddot{Q}_{ka}\dddot{Q}_{la}\rangle\,,
	\label{eq:dJ/dt}
\end{align}
where $\epsilon^{ikl}$ is the 3-dimensional Levi-Civita symbol, and the
angled brackets indicate a time average over several rotation periods.
Therefore the evolution of the angular momentum as function of time
simply reads
\begin{align}
	J_{\rm HMNS}(t)=J_{\rm merger}-t\left(\frac{dJ}{dt}\right)_{\rm cores}\,,
	\label{eq:Jtoy}
\end{align}
\ie the angular momentum decreases linearly with a slope constant over time.

In the second phase of the post-merger evolution, the radial oscillations of
the remnant star cores stop and we can approximate the system with a rotating
bar of fixed length. The frequency of rotation of the bar is $f_{\rm disk,2}$,
\ie the frequency towards which the oscillations of the previous phase tend to.
We assume the mass of the bar to be equal to $M_{\rm core}$, \ie the sum of the
two star cores, and denote its length by $L_{\rm bar}$. The bar is spinning
with constant angular velocity $\Omega_2=2\pi f_{\rm disk,2}$ around
the $z$-axis (see e.g., Sec. \ref{sec:lifetime} for additional
discussions on this assumption). The angular velocity $\Omega_2$ corresponds
to the value of $\Omega_{\rm disk}(t)$ in Eq.~\eqref{eq:omega} for late times.

It is then possible to compute the components of the quadrupole tensor of
the system yielding
\begin{subequations}
	\begin{align}
		\label{eq:quad.bar1}
		& Q^{\rm bar}_{xx}=\frac{M_{\rm core}
			L^{2}_{\rm bar}}{24}[1+\cos(2\Omega_2 t)] \\
		& Q^{\rm bar}_{yy}=\frac{M_{\rm core}
			L^{2}_{\rm bar}}{24}[1-\cos(2\Omega_2 t)] \\
		\label{eq:quad.bar2}
		& Q^{\rm bar}_{xy}=\frac{M_{\rm core}
			L^{2}_{\rm bar}}{24}\sin(2\Omega_2 t)=Q^{\rm bar}_{yx}\,.
	\end{align}
\end{subequations}
Proceeding analogously to the discussion above, we finally obtain the
expression of the angular momentum loss,
\begin{align}\label{eq: dJdt_bar}
	\left(\frac{dJ}{dt}\right)_{\rm bar}=\frac{2}
	{45}M^{2}_{\rm core}L^{4}_{\rm bar}\Omega^{5}_{2}\,.
\end{align}

Combining this result with that of the first phase according to
Eq.~\eqref{eq:dJ/dt}, we can express the evolution of the angular
momentum in the whole post-merger phase as a linear decrease in time with
a piecewise constant slope:
\begin{align}
	J_{\rm HMNS}(t)=
	\begin{cases}
		J_{\rm merger}-t\left(\frac{dJ}{dt}\right)_{\rm cores}
		&\text{ for }t\leq t_{\rm bar}\\
		J_{\rm bar}-t\left(\frac{dJ}{dt}\right)_{\rm bar}
		&\text{ for }t>t_{\rm bar}\\
	\end{cases}
	\label{eq:J}
\end{align}
where $t_{\rm bar}$ defines the time of change from the oscillating cores
phase to the bar phase, and $J_{\rm bar}=J_{\rm HMNS}(t_{\rm bar})$. The
evolution profile of the remnant angular momentum is illustrated in
Fig.~\ref{fig:model}.

The last missing piece of the model is a criterion which defines the time of
collapse of the merger remnant. Although a possible definition of this
quantity has already been proposed in the literature in relation to the maximum
bulk mass of the stable TOV solution \cite{Ciolfi2017} (see in particular Fig.
14 therein), in this work we will focus on the angular momentum evolution. In
fact, since HMNS are supported against collapse by differential rotation, the
loss of a sufficient fraction of the angular momentum present at merger should
trigger the collapse. In order to estimate the value of said fraction, we
looked at the 8 cases (out of the 42 models considered in \cite{Takami2015,
Rezzolla2016}) in which the HMNS collapsed during the simulation time, thereby
providing a value both for the time of collapse, $t_{\rm coll}$, and the
corresponding value of the angular momentum, $J_{\rm coll}$, for those
combinations of masses and EOS.

In order to make full use of our limited sample of points, it can be very
useful to analyze the ratio ${\epsilon_{\rm coll}=J_{\rm coll}/J_{\rm merger}}$
instead of $J_{\rm coll}$ alone.  If we consider $\epsilon_{\rm coll}$
depending exclusively on the dimensionless quantity $\tilde{M}=M_{\rm
NS}/M_{\rm TOV}$\footnote{In principle, $\epsilon_{\rm coll}$ should depend not
only on the total binary mass, but also on the mass ratio, the initial spins,
and most critically on the microphysics encoded in the EOS. Consistently with
the aims of this work, in order to develop a simple toy model we neglect here
the dependence on the mass ratio and spins. At leading order the dependence on
the EOS is taken into account by the dependence on the total mass of the
non-rotating NS, $M_{\rm TOV}$. Clearly, the present discussion is approximate
and valid only for heuristic purposes. We~plan to improve it in future work by
including dependencies on the quantities mentioned above.}, one can intuitively
imagine that for the most massive configurations, i.e., for $\tilde{M}=1$, the
collapse into a black hole (BH) happens promptly, so that the loss of angular
momentum is minimal, i.e.,  $\epsilon_{\rm coll}\approx1$ (see e.g.,
\cite{Agathos:2019sah} for additional discussions on the prompt BH formation
from BNSs). Similarly, for the less massive binaries, the post-merger remnant
can reach a stable configuration so that $\epsilon_{\rm coll}=0$.  As a
conservative estimate of the lower limit of the NSs masses, we select the
minimum mass of the observed NS mass distribution reported in
\cite{Kiziltan2013}, i.e., $M_{\rm min}\approx1.1$ $\msun$. Recalling that all
EOSs considered within this work have a TOV mass in the range $2.0-2.2~\msun$,
we obtain an approximate value for $\tilde{M}=M_{\rm min}/M_{\rm TOV}$ of 0.5 for $\epsilon_{\rm coll}=0$.
\begin{figure}[t]
	\centering
	\includegraphics[width=\columnwidth]{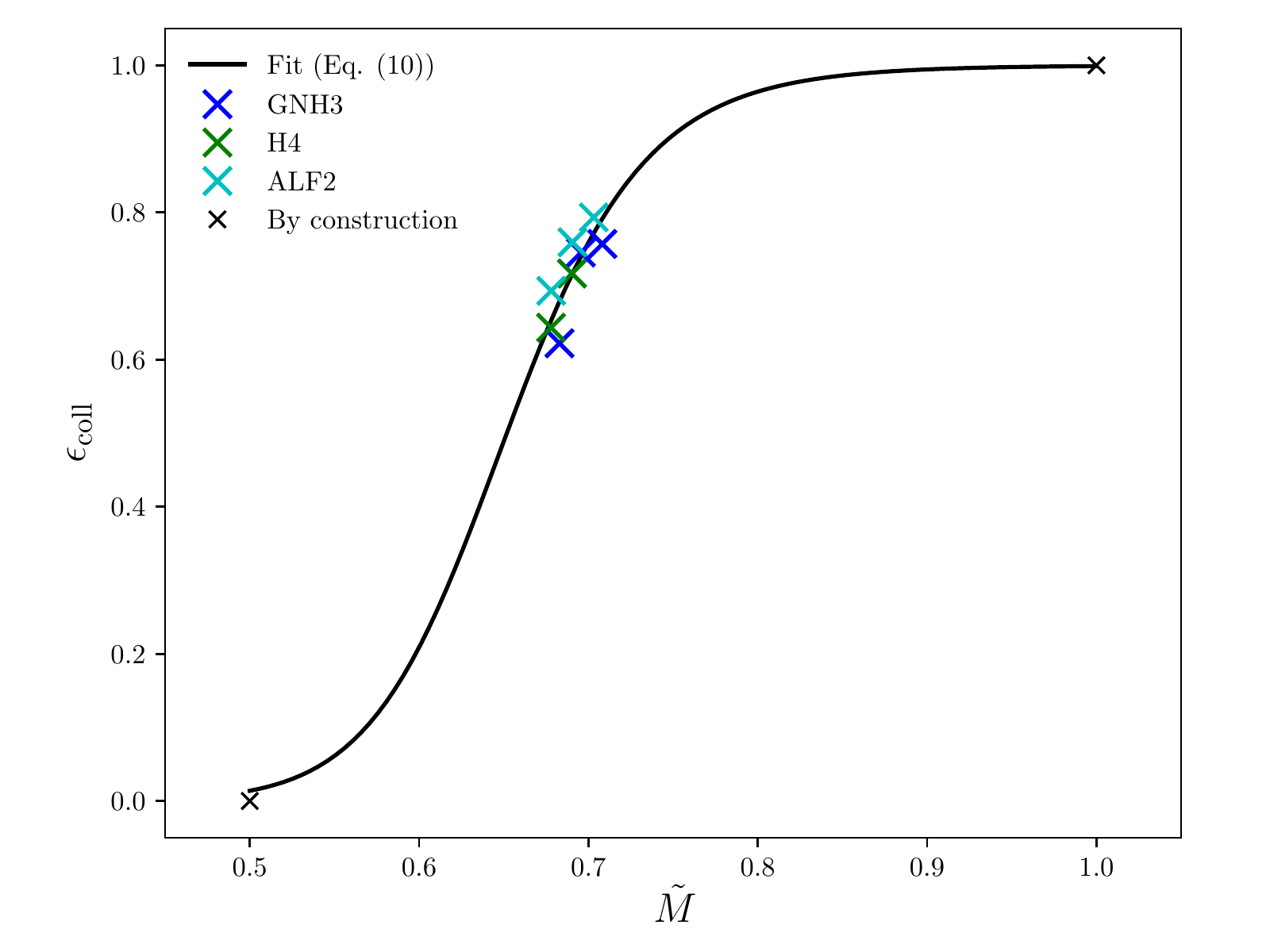}
	\caption{Ratio of the angular momenta at collapse and merger
		time, ${\epsilon_{\rm coll}=J_{\rm coll}/J_{\rm merger}}$, as a
		function of the dimensionless mass of the progenitor NSs,
		$\tilde{M}=M_{\rm NS}/M_{\rm TOV}$.}
	\label{fig:Jcoll}
\end{figure}

We show the complete sample of 10 points (8 data points plus the 2
limiting values) for $\epsilon_{\rm coll}$ as function of $\tilde{M}$ in
Fig. \ref{fig:Jcoll}. It can be seen that the data suggests a
$\tanh$-like behavior, which tends to 0 for low values of $\tilde{M}$,
has a step-like growth around a given value $b_1$ (with width $b_2$), and
converges to 1 for high values of $\tilde{M}$.  Concretely we can express
the expectations stated above with the following functional
relation\footnote{Note that there is considerable freedom in the choice
of the fitting function in Eq.~\eqref{eq:Jcoll}.  We have chosen the
fitting function requiring the least number of free parameters which
would not compromise accuracy. Furthermore, given the limited number of
points available within this work, the values of the parameters presented
here are at best indicative of the dependency trend of this function, and
are by no means meant to be definitive.  The very large region of data
space that our model is attempting to cover will only be better
understood with more extended simulations.}
\begin{equation}\label{eq:Jcoll}
	\epsilon_{\rm coll} =\frac{1}{2}\left[1+\tanh
	\left(\frac{\log_{10}\tilde{M}-\log_{10}b_1)}{b_2}\right)\right]\,.
\end{equation}
$b_i$ are two free parameters determined by fitting the data points and
which read ${b_1=0.6518\pm0.0081}$ and ${b_2=(5.4\pm1.2)\times10^{-2}}$.
The relation expressed in Eq.~\eqref{eq:Jcoll} is plotted in
Fig.~\ref{fig:Jcoll} together with the considered data and shows an
easily understandable physical meaning. Firstly, the extension of the
flat plateau at high masses (for $M_{\rm NS}>0.8 M_{\rm TOV}$), where the
fitting function is nearly equal to unity, can be interpreted as  a
formal definition of the correlation between prompt collapse and the mass
of the progenitor NSs. Conversely, if the progenitor NSs are less massive
than roughly 55\% of the TOV mass, the post-merger object can be
considered stable.

We remark, however, that the quality of the fit is substantially
mitigated by the lack of available data points. Testing the precise form of the behavior followed by the curve and the
derived conclusions will only be possible with future dedicated numerical
simulations.

\section{Testing the validity of the model}\label{sec:validity}
Now that the physical framework of our model has been defined, we turn
our attention to the numerical evaluation of its free parameters. For the
model presented in the preceding Sec.~\ref{sec:j_evol}, we need to find
for every BNS simulation the values of its parameters. After fixing the
value of the free parameters $a_i$ and $b_i$ in the phenomenological fit
functions \eqref{eq:Jmerger} and \eqref{eq:Jcoll}, respectively, to their
best-fit values, only the parameters of the toy model describing the
evolution of the angular momentum, \ie $M_{\rm core}$, $M_{\rm disk}$,
$R_{\rm disk}$, $k$, $b$, $v_0$, $t_{\rm bar}$ and $L_{\rm bar}$ (see
Eq.~\eqref{eq:J}), remain to be determined.

Assuming no mass loss during the inspiral and negligible mass loss in the
post-merger phase (the total value of the mass ejected in BNS merger
being of the order of $10^{-2}-10^{-3}$ $\msun$, see \eg
\cite{Bovard2017}), the sum of the mass of the disk and that of the cores
has to be equal to the initial total mass of the system, \ie $M_{\rm
	tot}=M_{\rm core}+M_{\rm disk}$, allowing us to write
\begin{align}
	\frac{M_{\rm core}}{M_{\rm tot}}
	+\frac{M_{\rm disk}}{M_{\rm tot}}=\mu_{\rm core}
	+\mu_{\rm disk}=1\,.
\end{align}
We can therefore consider a single parameter, e.g., $\mu_{\rm
	disk}~\in~[0,1]$, in place of both $M_{\rm core}$ and $M_{\rm disk}$.
Furthermore, we do not vary $R_{\rm disk}$ and $L_{\rm bar}$ directly but
rather work with their dimensionless equivalent ${\rho_{\rm disk}=R_{\rm
		disk}/R_{\rm NS}}$ and $\lambda_{\rm bar}=L_{\rm bar}/R_{\rm NS}$. This
reduces the set of parameters to 7: $\mu_{\rm disk}$, $\rho_{\rm disk}$,
$k$, $b$, $v_0$, $t_{\rm bar}$ and $\lambda_{\rm bar}$.

To define these parameters for each EOS, we employ a genetic algorithm
(GA). This method relies on the minimization of the difference between a
quantity predicted by the model, in our case the evolution of the angular
momentum, and the corresponding reference value, in our case the results
from the simulations of \cite{Takami2015, Rezzolla2016}.
\begin{figure*}[t]
	\centering
	\includegraphics[width=\columnwidth]{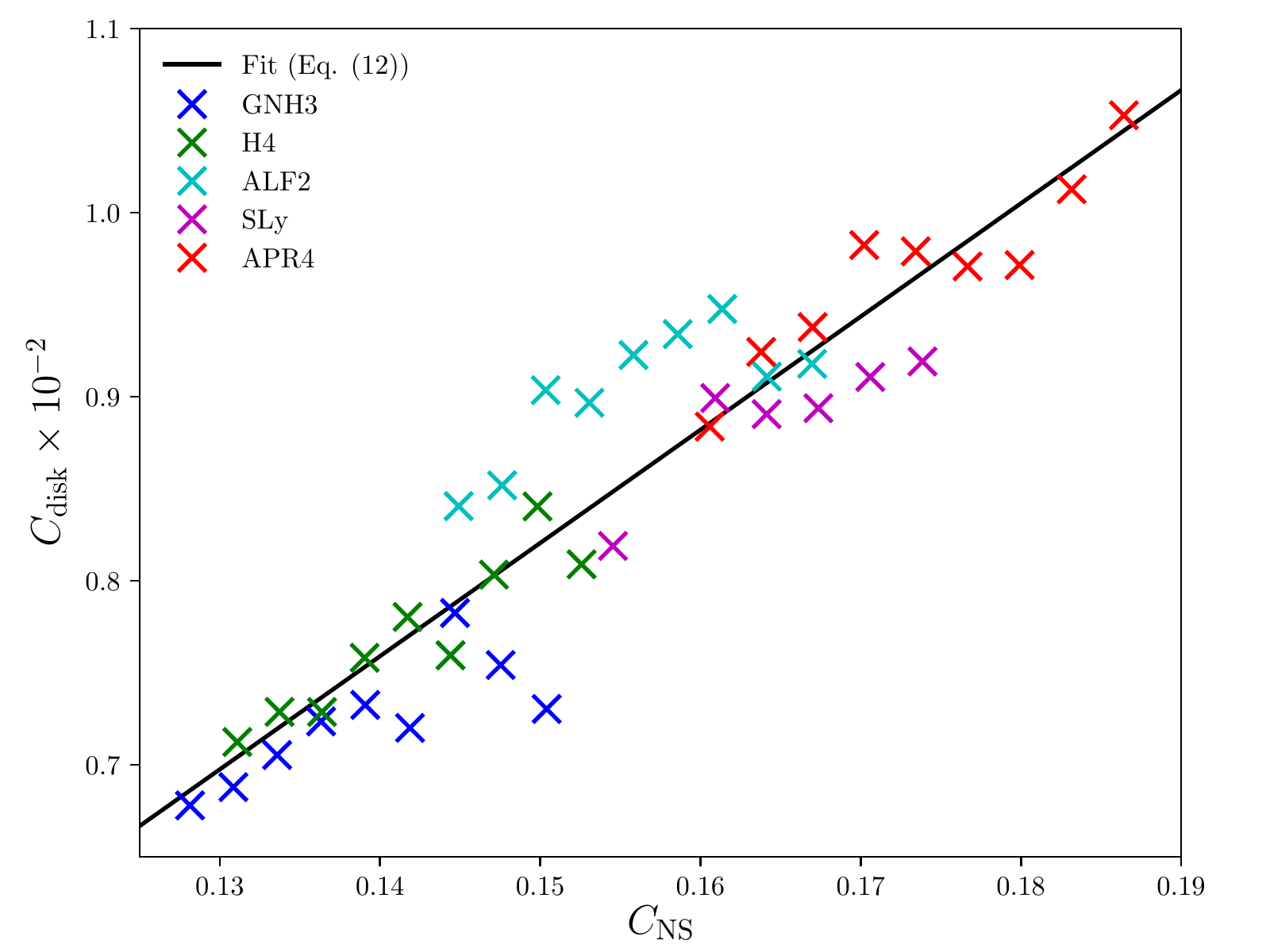}
	\includegraphics[width=\columnwidth]{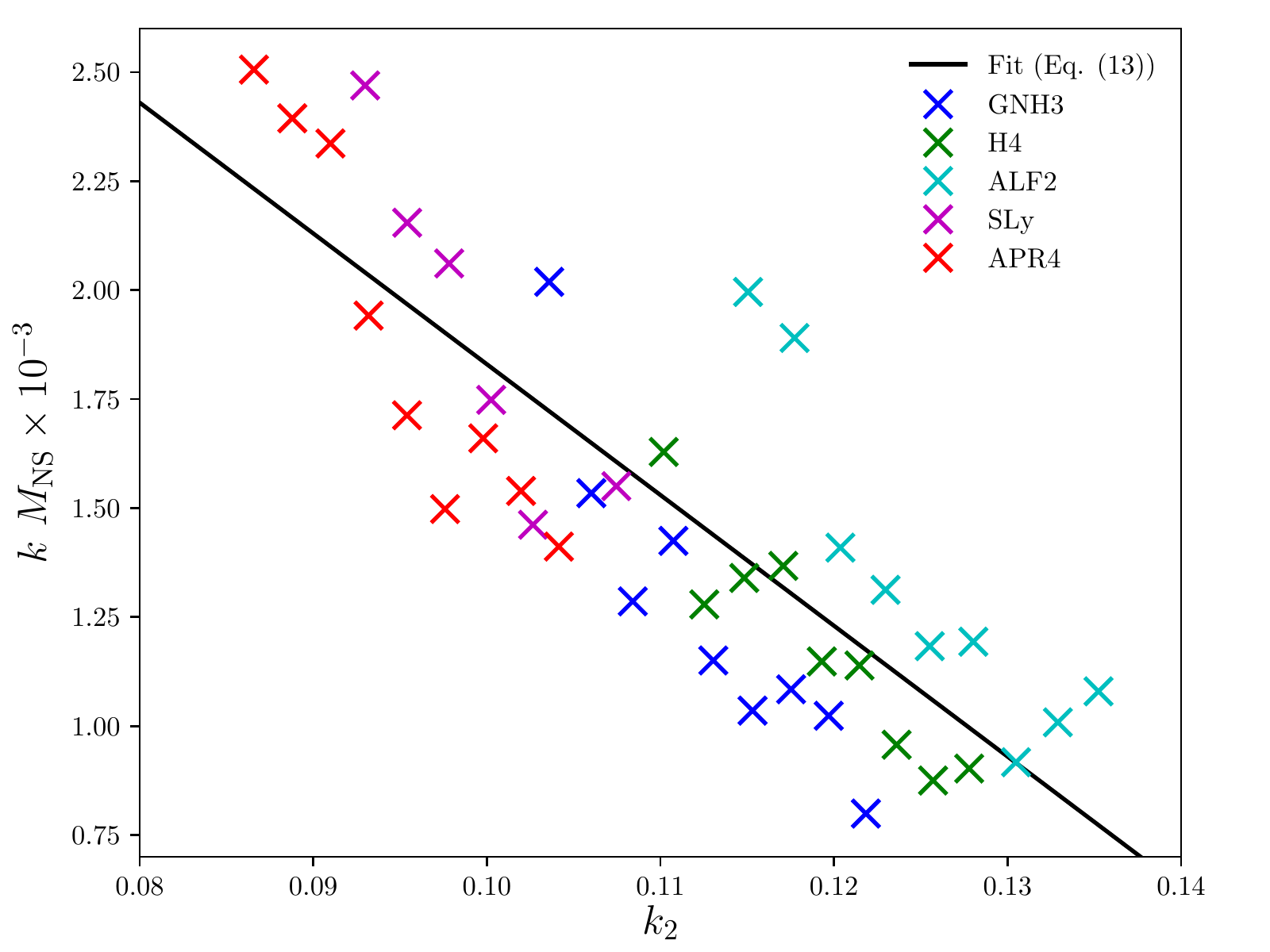}
	\caption{\textbf{Left panel}: Compactness of the disk $C_{\rm
			disk}$ as function of the initial compactness of the NSs $C_{\rm
			NS}$. Crosses are simulation data (color coded based on the model
		EOS), while the black line is the fit of Eq.~\eqref{eq:C_d}.
		\textbf{Right panel}: Dimensionless elastic constant $kM_{\rm
			NS}$ as function of the initial dimensionless Love number
		$k_{2}$. Crosses are simulation data (color coded based on the
		model EOS), while the black line is the fit of Eq.~\eqref{eq:k}.}
	\label{fig:Cd_CNS}
\end{figure*}

Furthermore, a GA also allows to set additional constraints that have to
be fulfilled in order for the algorithm to accept the fitness of a
parameter set. This feature becomes particularly useful in our analysis.
In fact, by fitting the evolution of the angular momentum with
Eq.~\eqref{eq:J} alone, we would have more free parameters than degrees
of freedom (DOFs): in Eq. \eqref{eq:J} we have three DOFs ($(dJ/dt)_{\rm
	cores}$, $J_{\rm bar}$ and $(dJ/dt)_{\rm bar}$) while in the previous
paragraph we listed 7 free parameters. Therefore, in order be able to
uniquely define each parameter of the model, we need to introduce an
additional independent prediction of the model that we can test against the
reference at the same time. We choose these constraints to be the values
$f_1$ and $f_2$ of the PSD.

For a more formal definition of the optimization problem, its application
to our analysis and the table of the resulting best-fits, Tab. \ref{tab:
GA_results}, see App. \ref{Sec: Appendix A}. There, we also show how the
dimensionless disk radius $\rho_{\rm disk}$ and the damping constant $b$
can be fixed to the universal values of $3.5$ and $0.005$, respectively.
In this way, one can effectively reduce the number of free parameters to
5: $\mu_{\rm disk}$, $k$, $v_0$, $t_{\rm bar}$ and $\lambda_{\rm bar}$.

With both the mathematical and the numerical setup ready, we can apply
our procedure to all 42 cases considered within this work. Overall, the
mean value of $\rchitwo$ (defined as in Eq. \eqref{eq:minfun}) that we
achieve across the models is $\approx$ 1.44 $\times 10^{-4}$, while the
frequencies $f_1$ and $f_2$ are recovered with an approximate deviation
from the values of \cite{Takami2015, Rezzolla2016} of 3.1\% and 2.8\%,
respectively.

This level of accuracy is \textit{per se} already remarkable. It
systematically shows for the first time that a mechanical toy model such
as the one described in Sec. \ref{sec:j_evol} can in principle precisely
predict the evolution of the angular momentum during the whole
post-merger evolution together with other fundamental spectral
quantities. In the following paragraphs we further investigate the
solidity of the results with different approaches.

\subsection{Universal behavior of the GA results}
As a first test of the validity of the GA results, one can look for
universal behaviors of the fitted quantities. In fact, these are expected
since post-merger quantities, such as the compactness of the cores and
the elastic constant, can intuitively be related to characteristics of
the progenitor NSs, such as the initial compactness and deformability,
respectively. Eventual
positive relations found using the GA best-fit values would be a very strong
confirmation of the fact that the GA results are not just sourced by
degeneracies in the fitting procedure, but have a deeper physical meaning.

The first example we consider is the already mentioned case of the
correlation between the compactness of the cores and of the progenitor
NSs. Recalling that the radius of the disk has been fixed at the value of
3.5$R_{\rm NS}$ and using the values of $\mu_{\rm disk}$ resulting from
the GA, it is possible to compute the disk compactness $C_{\rm disk}$. We
find that this can be expressed as function of the initial compactness of
the NSs by the linear relation
\begin{align}\label{eq:C_d}
	C_{\rm disk}=c_1 C_{\rm NS}+c_2\,,
\end{align}
with the free parameters ${c_1=(6.15\pm 0.37)\times 10^{-2}}$
and~${c_2=-(1.02\pm 0.34)\times 10^{-3}}$. Eq.~\eqref{eq:C_d} is shown in
the left panel of Fig.~\ref{fig:Cd_CNS} together with the GA best-fits.
Using this relation, $\mu_{\rm disk}$ becomes a function of $C_{\rm
	disk}$ and $R_{\rm disk}$ and the sole knowledge of $M_{\rm NS}$ and
$R_{\rm NS}$ determines the values of $\mu_{\rm disk}$, $\mu_{\rm core}$
and $\rho_{\rm disk}$.

In the right panel of Fig.~\ref{fig:Cd_CNS}, we show the values of the
dimensionless elastic constants $kM_{\rm NS}$ as a function of the
$\ell=2$ dimensionless tidal Love number $k_{2}$ at infinite separation
(see e.g., \cite{Rezzolla2016} for more details). Not surprisingly, the
parameter that controls the strength of the interaction between the stars
in our model is actually correlated with their tidal deformability. We
find that the data can in fact be fitted with the following relation
\begin{align}
	kM_{\rm NS}=d_1 k_{2}+d_2\,,
	\label{eq:k}
\end{align}
with the free parameters ${d_1=-(2.96\pm 0.30)\times 10^{-2}}$ and
${d_2=(4.76\pm 0.34)\times 10^{-3}}$.

As a remark, note that our goal in finding universal relations 
linking the fitted quantities to the initial conditions of the system 
is only to validate the robustness of the model and of the GA results. 
They are not made part of the post-merger model in order to reduce its 
number of free parameters. We leave such a development for a future 
investigation.

For $v_0$, $t_{\rm bar}$ and $\lambda_{\rm bar}$ a clear universal relation
could not be found. However, even these parameters can be used to
support the validity of our model. Indeed, taking as values the average of the GA
results over all configurations, \ie $v_0=0.115$, $t_{\rm bar}=4.83$ ms and
$\lambda_{\rm bar}=4.18$, some useful consistency checks can be proposed. For
instance, the mean value for $t_{\rm bar}$ corresponds to the expectation
stated in \cite{Takami2015} regarding the validity of the first phase of the
toy model, \ie that it models the first $\sim 5$ ms of the post-merger.

In conclusion, therefore, we can assume that the GA best-fits are
consistent with the expected universal behaviors and truly reflect
physical quantities.

\subsection{The representative case of \texttt{GNH3-M1300}}
\begin{figure*}[t]
	\includegraphics[width=\textwidth]{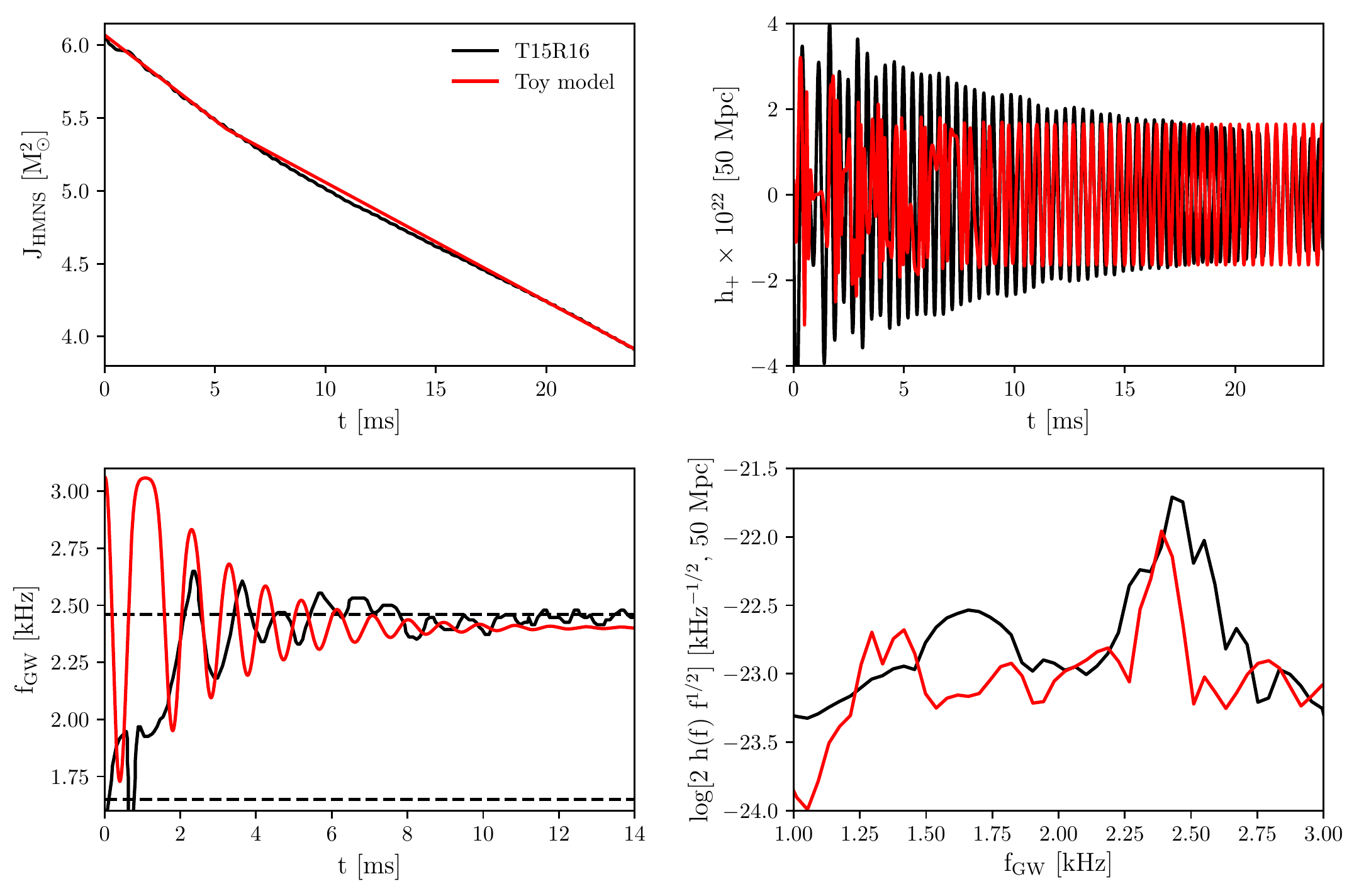}
	\caption{Comparison between (top to bottom, left to right) the
	post-merger evolution of the angular momentum, the frequency of the GW
	signal, the GW plus polarization amplitude $h_+$ for a source
	at 50 Mpc and the PSD predicted by our model (red lines) and as
	computed in the simulations of \cite{Takami2015, Rezzolla2016} (black
	lines, here labeled T15R16 for brevity) for the representative case of
	\texttt{GNH3-M1300}. The dotted lines in the second panel represent the
	values of $f_1$ and $f_2$ as defined in \cite{Rezzolla2016}.}
	\label{fig:comp_GNH3_1300}
\end{figure*}

In this following discussion, we also want to make sure that our model,
together with its best fit parameters, actually has a predictive power
and that the latter is not just limited to the angular momentum
evolution. Thus, in order to test its accuracy, we focus our
attention on the representative case of the simulation
\texttt{GNH3-M1300}, shown in Fig.~\ref{fig:comp_GNH3_1300}. The choice
of this particular case is driven by the analysis of the same
configuration made in Appendix~A of \cite{Takami2015}. Specifically, we
want to propose a direct comparison of our results to the ones shown in
Figs.~18 and 19 of the reference.

In Fig.~\ref{fig:comp_GNH3_1300} we show the post-merger evolution of the
angular momentum in the top left panel, and the frequency of the GWs emitted in
the post merger phase as predicted by our model in the bottom left
panel. In all panels of Fig.~\ref{fig:comp_GNH3_1300}, the black lines
represent the results from the simulation by \cite{Takami2015, Rezzolla2016},
while the red lines correspond to the results of our model. Furthermore, the
dotted lines in the second panel represent the values of $f_1$ and $f_2$ as
defined in \cite{Rezzolla2016}.

In both panels, the overlap between the model prediction and the reference
is remarkable. This is even more true comparing $f_{\rm GW}$ to the same
quantity in Fig. 18 of \cite{Takami2015}. The fact alone that such
accuracy is possible is non-trivial and should be considered as a strong
evidence for the solidity of the toy model.

In order to further test of the robustness of our model, we compute the GW
polarization amplitude $h_+$ and the PSD of the GW amplitude as defined in
\cite{Takami2015}. The overlap between the model prediction and reference is
plotted in the top and bottom right panels of Fig.~\ref{fig:comp_GNH3_1300},
respectively. The similarity between the curves is less evident than in the
previous cases although still striking, particularly when considering the
simplicity of the present model.

These results clearly show that the toy model and the fitting procedure
employed within this work to describe the HMNS angular momentum evolution
are already very accurate in the prediction of several fundamental
quantities over the time range explored by the reference simulations.
However, at the same time Fig.~\ref{fig:comp_GNH3_1300} also points to
the fact that our analysis might still require additional refinements, as
for instance the inclusion of more reference quantities in the GA, which
could improve the precision of the derived spectral features. This
possibility will be addressed in future work.

\subsection{The HMNS lifetime \label{sec:lifetime}}
Since the model accurately reproduces the HMNS evolution within the first
25 ms after the merger, it can be very useful to investigate its validity
in terms of the HMNS lifetime. To do so, we calculate the collapse time
of all 42 configurations using a linear combination of Eqs.~\eqref{eq:J}
and \eqref{eq:Jcoll}, i.e., with
\begin{align}\label{eq: t_coll}
	t_{\rm coll}=t_{\rm bar}+\frac{J_{\rm bar}-J_{\rm coll}}{(dJ/dt)_{\rm bar}}
\end{align}
where $J_{\rm bar}=J_{\rm merger}-t_{\rm bar}(dJ/dt)_{\rm toy}$. We then
compare the model predictions to the quasi-universal relation found in
\cite{Lucca:2019ohp} connecting the HMNS lifetime to the mass of the
progenitor NSs.

\begin{figure}[t]
	\centering
	\includegraphics[width=\columnwidth]{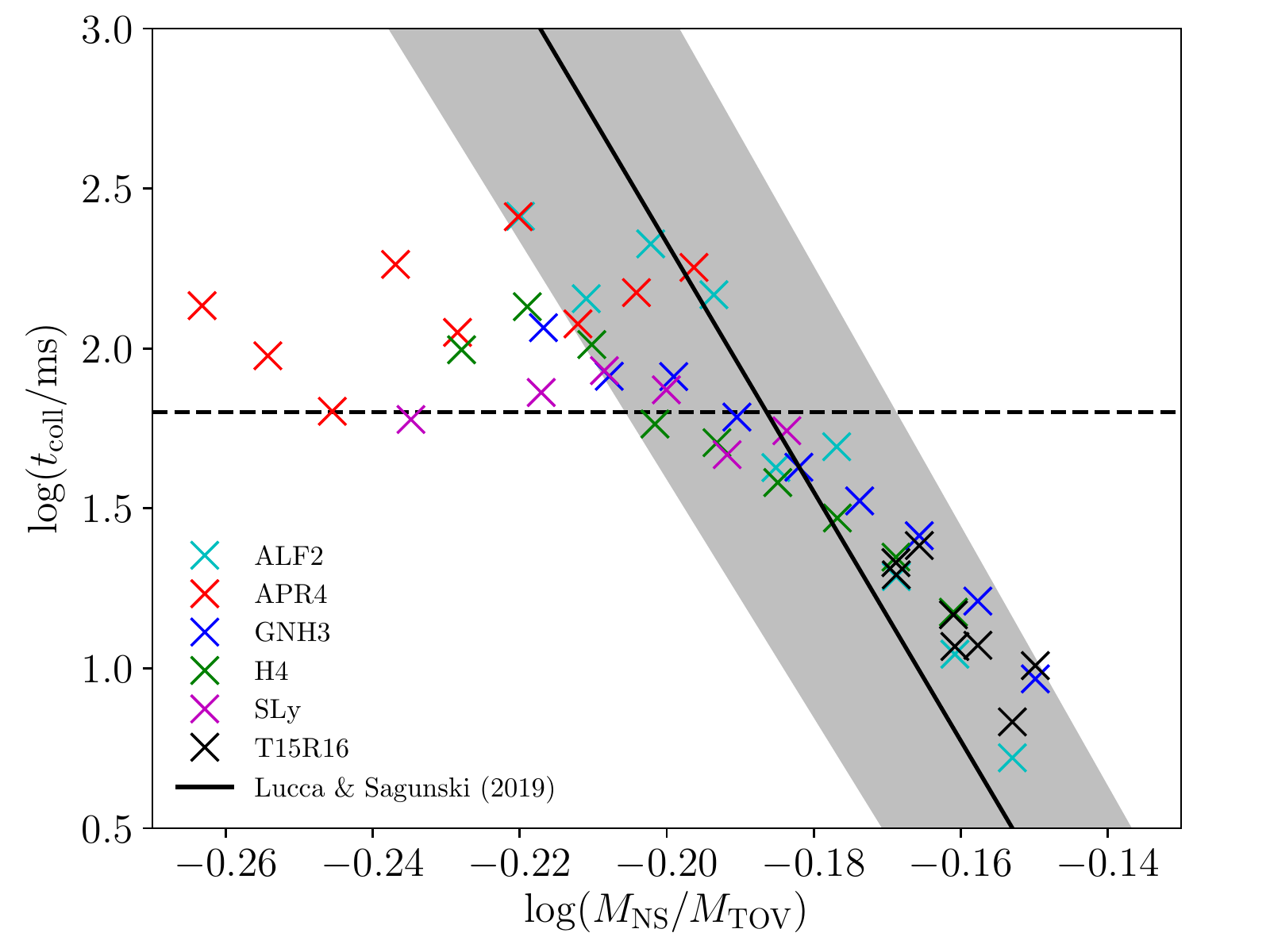}
	\caption{Comparison between the predictions for the HMNS
	lifetimes calculated using our toy model (colored crosses) and as
	given in the reference simulations by \cite{Takami2015,
	Rezzolla2016} (black crosses, again labeled T15R16 for
	shortness), and the relation found in \cite{Lucca:2019ohp} (black
	line with additional 1$\sigma$ uncertainty region in gray). The
	horizontal dotted line corresponds to 60 ms.}
	\label{fig: lifetime}
\end{figure}

In Fig.~\ref{fig: lifetime}, we show the relation\footnote{Here we neglect the
dependence on the mass-ratio $q$ since all considered configuration are equal
mass binaries.} of \cite{Lucca:2019ohp} as a black line together with the
1$\sigma$ uncertainty region in gray.  The black crosses represent the collapse
times of the 8 reference configurations analyzed in \cite{Takami2015,
Rezzolla2016} that collapsed within the simulation time. Although not
explicitly shown in the figure, these data points come with intrinsic
uncertainties due to known numerical issues such as the chosen resolution (see
e.g., \cite{Hotokezaka2013c, Rezzolla2016, Bauswein2010Testing} for more
details). However, as these errors are particularly difficult to estimate and
are not discussed in details in \cite{Takami2015, Rezzolla2016}, only the
single data points are displayed in the figure. The colored crosses represent
the model predictions for the different~EOSs.

In the figure it is possible to observe that, although for low values of
the collapse times the model predictions are very accurate, they start to
diverge from the fit at approximately 60 ms (represented by the
horizontal dotted line). There are two main possible explanations for
this effect: either Eq.~\eqref{eq:Jcoll} is incorrect in the low mass
region (or equivalently for long-living configurations) or there is a
late-time softening of the angular momentum decrease, which is not
captured by the toy model as it is defined in Sec. \ref{sec:j_evol}.

We can test the first possibility by setting $J_{\rm coll}$ to zero in
Eq.~\eqref{eq: t_coll}, i.e., maximizing the collapse time for all
configurations. As it turns out, in this case one only obtains an overall
shift of the more massive configurations towards higher collapse times,
leaving effectively unchanged the predictions for low-mass binaries.

This points towards the idea that the toy model is currently missing some
late-time feature of the angular momentum evolution. Although a detailed
modeling of this late-time dynamics will only be possible with the help
of extended numerical simulations, in a first approximation one could
introduce a third late-time phase in the angular momentum evolution with
a reduced emission of~GWs.

If we assume that the shape of the HMNS in this phase is still bar-like and
that no significant amount of mass is being ejected from the two cores (see
e.g., Tab.~1 of~\cite{Shibata2019Merger} for a quantitative overview),
according to Eq.~\eqref{eq: dJdt_bar} there are only two parameters effectively
affecting the angular momentum loss rate: the length and the rotation frequency
of the bar. As the $f_2$ peak in the PSD can often be very sharp (see e.g.,
Fig. 6 of \cite{Rezzolla2016}), we can assume in this approximation that the
rotation frequency of the HMNS, which is directly proportional to the frequency
of the GW signal, is nearly constant after the transient phase. This
conclusion can also be justified in light of recent long-lasting numerical
simulations such as the ones performed in \cite{Ciolfi:2019fie} (see in
particular Fig. 16 therein). This leads to the idea that only the length of
the bar may in fact vary over long time ranges.

This fact can be intuitively explained by recalling that as more
angular momentum is lost, the rotation velocity of the HMNS becomes less
efficient in supporting the star against the gravitational force pulling
matter toward the inside. Thus, it leads to an overall compression of the
HMNS and therefore a shrinking of its size.

In order to address this limitation of the toy model, we tested straightforward
options such as the introduction of a third phase sharply setting on at a time
$t_3$ and with a reduced value of $L_{\rm bar}$ or a transformation of type
${L_{\rm bar}\to L_{\rm bar}-At}$ in Eqs.
\eqref{eq:quad.bar1}-\eqref{eq:quad.bar2}, where $A$ is a constant driving the
rate of change of the bar's length. While such a modification of the toy model
presented here is ultimately desirable to properly capture the late-time
evolution of the angular momentum loss, the aforementioned attempts to address
it could not fully solve the problem yet. We therefore leave a more in-depth
analysis of this aspect to future work, noting also that a proper and
systematic exploration of the late-time dynamics of the merger remnant needs
data from numerical simulations extending to $\sim 100$ ms after merger
(see e.g., \cite{Ciolfi:2019fie, Ciolfi:2020hgg} for related
discussions).

\section{Conclusions}\label{sec:conclusions}
In the last decade, with the help of a number of fully general
relativistic numerical simulations, we achieved an unprecedented
understanding of the post-merger evolution of binary NSs. Moreover, the
main features of the picture emerging from these simulations have been
recently confirmed by the worldwide multimessenger observation of a NS
merger, GW170817.

However, the rising complexity necessary to numerically describe these
events increases the difficulty in interpreting the physics underlying
them. Examples of currently debated aspects of the post-merger evolution
involve, for instance, the source of the secondary peaks in the power
spectral density and the HMNS lifetime. It is therefore often helpful to
simplify the dynamics of the system by means of toy models. In this
regard, a particularly useful quantity to track over the whole binary
evolution is the angular momentum of the post-merger object. This is
because this parameter is tightly connected to key observables, such as
the GW emission, and it is also a fundamental ingredient to understand
the final stages of the binary evolution, as the star is held against
collapse by its high rotational velocity.

In this work, in order to construct a viable toy model for the angular
momentum evolution of the HMNS, we first base ourselves on the numerical
results of \cite{Takami2015, Rezzolla2016} to find a universal relation
describing the angular momentum at the time of merger. Then, in order to
evaluate the transient phase we employ the toy model defined in~\cite{Takami2015}. Subsequently, we extend this model to later times by
approximating the HMNS as a rotating bar of fixed length. To complete our
mathematical setup, we find a second relation connecting the angular
momentum of the HMNSs, this time at collapse time, to the initial
parameters of the system. For this last step we use once again the
results of \cite{Takami2015, Rezzolla2016}.

Once our extended toy model is defined, to determine the free
parameters needed to complete the equations we use a GA. In order to perform a more realistic evaluation of the model, we
also include constraints from the evolution of the GW frequency in the
GA, which can also be predicted by means of our toy model.

Despite the simplified mathematical setup, several interesting
conclusions can be drawn from our results. First of all, we present two
original relations (one universal and one quasi-universal) describing the
angular momentum at critical moments in the binary evolution, the merger
and the collapse.

Furthermore, we systematically show that it is in principle
possible to precisely predict the evolution of the angular momentum of 
a BNS merger remnant as well as many characteristics of its GW signal
by means of the aforementioned toy model. Note that, although other models 
of the post-merger GW signal are available in the literature, such
as the one developed by \cite{Bernuzzi:2015rla, Breschi20}, here we follow a
different approach. For instance, we do not assume a functional form for the GW
signal (which in principle could be arbitrary and not linked to the properties
of the merger remnant) and then fit its parameters on numerical data. Instead, we start
from a mechanical model of the internal structure of the remnant, which has the advantage of allowing a tighter and more physical connection between the model's free parameters and the derived quantities. Moreover, the
primary goal of our model is to recover the angular momentum evolution, and not the
evolution of the GW signal itself, which in our case only becomes a derived quantity.

Given our setup, the fact that extremely
accurate fits are possible for every EOS and mass configuration already
supports the validity of our model. However, we further investigate the
precision and the solidity of our results by suggesting intuitive universal
relations that the GA results follow nicely.  Moreover, using the same fitting
parameters we also derive additional spectral properties, such as the
polarization amplitudes and the PSD, and compare them to the reference
simulations by \cite{Takami2015, Rezzolla2016} for a representative case.

Additionally, we compute the HMNS lifetime for every mass configuration
considered within this work. Using the results of \cite{Lucca:2019ohp},
we then show that our model is very predictive up to lifetimes roughly in
the order of ${60}$~ms, and that there must be a late-time
phase in the binary evolution, where the length of the spinning bar is
significantly reduced and the loss of angular momentum is even slower
than in the quasi-stationary phase.

Finally, however, it is important to underline some limitations of our
analysis.  First of all, we limited our initial sample to equal-mass
binaries and future work will be needed in order to enlarge the
progenitor NS mass-ratio range. Secondly, the amount of EOSs considered
within this work should also be increased. Other constraints come from
the simulations used to produce the initial data. There, effects like
fluid viscosity, electromagnetic fields and neutrino transport have been
neglected. Also, thermal effects in the fluid are accounted for only
approximately via an ideal-gas EOS contribution. Furthermore, we will
devote future work to extend the number of constraints as well as the set
of minimized functions, e.g., by including the evolution of the energy
emitted through GWs.

\section*{Acknowledgments}
\label{sec:acknowledgements}
The authors thank T. Hambye, P. Mertsch, and \linebreak L. Rezzolla for their numerous
useful comments and inputs. Furthermore, it is a pleasure to thank K. Takami
for providing access to the simulation data on which the present work is based.
ML is supported by the ``Probing dark matter with neutrinos" ULB-ARC
convention and by the IISN convention 4.4503.15. LS has been supported by the 
Deutsche Forschungsgemeinschaft (DFG)
through the Emmy Noether Grant No. KA 4662/1-1. FG is supported by the European
Research Council grant EUROPIUM (grant No.  677912). CMF is supported by the ERC Synergy Grant ``BlackHoleCam - Imaging the
Event Horizon of Black Holes'' (Grant 610058).

\appendix
\section{The setup of the genetic algorithm}\label{Sec: Appendix A}
To find the desired relation linking the evolution of the angular
momentum and the initial parameters of the system we set up a fitting
procedure as a constrained non-linear optimization problem:
\begin{equation*}
	\begin{array}{ll@{}ll}
		\text{minimize}  & \displaystyle f(\vec{x}) &\\
		\text{subject to}& \displaystyle g_{j}(\vec{x})\leq 0, &   &
		j=1 ,..., n\\
		& \displaystyle x_{L,i}\leq x_{i}\leq x_{R,i},
		& &i=1 ,..., 7\,,
	\end{array}
\end{equation*}
where $f(\vec{x})$ is the objective function (\ie the function whose
minimum corresponds to the optimal parameters' values), $g_j(\vec{x})$
are a set of constraints, $\vec{x}$ is a 7-dimensional vector whose
components are the model parameters, \eg $\vec{x}=\left(\mu_{\rm disk},
\rho_{\rm disk}, k, b, v_0, t_{\rm bar}, \lambda_{\rm bar}\right)^{T}$,
and $\{x_{L,i}$, $x_{R,i}\}$ are lower and upper bounds for the
parameters.

As the minimization function $f(\vec{x})$ we choose the least squares
$\tilde{\chi}^2$:
\begin{align}
	\tilde{\chi}^2=\frac{1}{l}\sum_{\rm k=1}^{l}\frac{(J_{\rm HMNS}(t_{\rm k})
		-J_{\rm data}(t_{\rm k}))^2}{J_{\rm data}(t_{\rm k})}\,,
	\label{eq:minfun}
\end{align}
where $J_{\rm HMNS}$ is the angular momentum value predicted by
Eq.~\eqref{eq:J} and $J_{\rm data}$ is the value provided by the
simulation data (see Fig.~\ref{fig:j_example}). As summarized in
Fig.~\ref{fig:model}, the decrease in $J_{\rm HMNS}(t)$ as predicted by
our model consists simply in two linear pieces of constant slope and it
is therefore very unlikely that it can accurately describe the transition
phase from the oscillating-spheres configuration to the rotating bar one.
Acknowledging this limitation we include deviations of the model from the
data in the computation of $\rchitwo$ only over two specific time
windows: the first $3$ ms immediately after merger; and the $6$ ms
before collapse, or before $t=25$ ms for the simulations in which the
collapse was not observed during the simulation time.  Therefore the
index $k$ in Eq.~\eqref{eq:minfun} effectively only runs over the data
points within these two time windows.

We include furthermore two constraints based on the GW frequencies $f_1$
and $f_2$:
\begin{align}
	g_1&=\frac{f_{\rm 1,ref}-f_{\rm 1,model}(\vec{x})}{f_{\rm 1,ref}}
	-0.05f_{\rm 1,ref}\\
	g_2&=\frac{f_{\rm 2,ref}-f_{\rm 2,model}(\vec{x})}{f_{\rm 2,ref}}
	-0.05f_{\rm 2,ref}\,,
\end{align}
where $f_{\rm 1,model}$ and $f_{\rm 2,model}$ are the minimum frequency
of the post-merger GW signal and its limit frequency at late times,
respectively, as predicted by our model, and $f_{\rm 1,ref}$ and $f_{\rm
	2,ref}$ are the corresponding reference values according to the
simulations of \cite{Takami2015, Rezzolla2016}. The constraint functions
are designed to allow for a tolerance of 5\% on the deviation of the
model frequencies from the reference ones.

An optimization method based on a gradient search would most likely
converge to a local minimum of the problem, rather than to the global
one. To avoid this behavior the gradient of the 7-dimensional parameter
space has to be mapped out with very high precision, which is
computationally very expensive and impractical. We have therefore elected
to apply a non-gradient based search algorithm. Among this class of
algorithms we select a genetic algorithm (GA), as it is known to be very
efficient in searching for global minima. The GA we employ is
provided by the Python- based package \texttt{pyOpt}~\cite{perez2012}, of
which we make use of the optimizer \texttt{NSGA2} (Non Sorting Genetic
Algorithm II) \cite{deb2002}.

Over the course of the generations produced by the GA a given parameter
evolves towards its optimal value. Some parameters however have a greater
influence on the value of the fitness function then others. Furthermore
we observe that some parameters remain nearly constant during the
optimization process (\ie they converge nearly immediately to the optimal
value), and that the small variations they experience do not affect
significantly the value of $\rchitwo$. We take advantage of this
behavior to further reduce the number of parameters.
\begin{table}[t]
	\centering
	\scriptsize
	\begin{tabular}{|c|c|c|c|c|c|}
		\hline
		EOS & $\displaystyle \frac{\mu_\textrm{disk}}{10^{-2}}$ &
		$\displaystyle \frac{k}{10^{-4}}$ & $\displaystyle
		\frac{v_0}{10^{-1}}$ & $\displaystyle t_\textrm{bar}$ &
		$\lambda_\textrm{bar}$ \\
		\hline
		ALF2-1200 & 10.1 & 9.00 & 1.12 & 7.38 & 3.52 \\
		ALF2-1225 & 10.1 & 8.23 & 0.77 & 5.00 & 4.14 \\
		ALF2-1250 & 10.5 & 7.34 & 0.43 & 5.32 & 3.80 \\
		ALF2-1275 & 10.3 & 9.36 & 0.84 & 6.83 & 3.42 \\
		ALF2-1300 & 10.4 & 9.10 & 0.59 & 4.24 & 4.73 \\
		ALF2-1325 & 10.3 & 9.91 & 0.60 & 6.30 & 3.85 \\
		ALF2-1350 & 10.3 & 10.4 & 0.84 & 5.12 & 4.64 \\
		ALF2-1375 & 9.71 & 13.7 & 1.56 & 4.50 & 4.31 \\
		ALF2-1400 & 9.62 & 14.3 & 1.62 & 3.97 & 4.74 \\
		\hline
		APR4-1200 & 9.63 & 11.8 & 1.13 & 7.37 & 4.28 \\
		APR4-1225 & 9.88 & 12.6 & 1.63 & 6.11 & 4.14 \\
		APR4-1250 & 9.83 & 13.3 & 1.59 & 4.11 & 4.43 \\
		APR4-1275 & 10.1 & 11.7 & 0.65 & 5.46 & 3.77 \\
		APR4-1300 & 9.88 & 13.2 & 0.55 & 4.25 & 3.94 \\
		APR4-1325 & 9.62 & 14.7 & 0.56 & 4.18 & 2.90 \\
		APR4-1350 & 9.45 & 17.3 & 0.62 & 3.79 & 3.16 \\
		APR4-1375 & 9.67 & 17.4 & 0.56 & 3.86 & 2.92 \\
		APR4-1400 & 9.88 & 17.9 & 0.50 & 3.99 & 2.70 \\
		\hline
		GNH3-1200 & 9.26 & 6.66 & 1.34 & 2.29 & 5.34 \\
		GNH3-1225 & 9.20 & 8.36 & 1.61 & 1.48 & 5.00 \\
		GNH3-1250 & 9.24 & 8.67 & 1.60 & 6.10 & 4.70 \\
		GNH3-1275 & 9.29 & 8.12 & 1.15 & 2.18 & 5.02 \\
		GNH3-1300 & 9.22 & 8.85 & 1.09 & 5.48 & 4.97 \\
		GNH3-1325 & 8.88 & 10.8 & 1.63 & 1.59 & 4.64 \\
		GNH3-1350 & 9.46 & 9.53 & 1.46 & 4.51 & 5.02 \\
		GNH3-1375 & 8.95 & 11.2 & 1.53 & 6.39 & 4.60 \\
		GNH3-1400 & 8.50 & 14.4 & 1.89 & 6.08 & 4.38 \\
		\hline
		H4-1200 & 9.51 & 7.52 & 1.59 & 2.51 & 5.34 \\
		H4-1225 & 9.54 & 7.14 & 1.13 & 3.19 & 4.83 \\
		H4-1250 & 9.35 & 7.66 & 1.09 & 4.92 & 4.75 \\
		H4-1275 & 9.54 & 8.93 & 1.53 & 4.68 & 5.01 \\
		H4-1300 & 9.63 & 8.83 & 1.05 & 2.60 & 4.96 \\
		H4-1325 & 9.20 & 10.3 & 1.35 & 6.45 & 4.62 \\
		H4-1350 & 9.56 & 9.92 & 1.08 & 5.84 & 4.85 \\
		H4-1375 & 9.81 & 9.30 & 1.02 & 6.24 & 5.09 \\
		H4-1400 & 9.28 & 11.6 & 1.32 & 6.11 & 4.60 \\
		\hline
		SLy-1200 & 9.27 & 12.9 & 2.01 & 7.38 & 4.78 \\
		SLy-1250 & 9.78 & 11.7 & 1.08 & 7.36 & 4.45 \\
		SLy-1275 & 9.49 & 13.7 & 1.58 & 6.03 & 3.83 \\
		SLy-1300 & 9.34 & 15.9 & 1.79 & 5.68 & 3.55 \\
		SLy-1325 & 9.34 & 16.3 & 1.42 & 3.17 & 3.97 \\
		SLy-1350 & 9.25 & 18.3 & 1.39 & 5.25 & 3.29 \\
		\hline
	\end{tabular}
	\caption{
	Genetic algorithm results for the set of 5 parameters.
	The various columns denote the EOS and the average gravitational
	mass $M_\textrm{NS}$ at infinite separation in $M_\odot/10^{-3}$
	("EOS-M$_\textrm{NS}$"), the dimensionless parameter
	$\mu_\textrm{disk}=M_\textrm{disk}/M_\textrm{tot}$, the elastic
	constant $k$ in $M_{\odot}^{-1}$, the speed at plunge $v_0$, the
	time $t_\textrm{bar}$ in ms at which the system switches from the
	oscillating configuration to the non-oscillating one, and the
	dimensionless factor
	$\lambda_\textrm{bar}=L_\textrm{bar}/R_\textrm{NS}$.
	}
	\label{tab: GA_results}
\end{table}

To quantify how much a given parameter $X$ changes along the generations,
we define the quantity ${\Theta=\Delta X/X_{\rm opt}}$, where $\Delta
X=|X-X_{\rm opt}|$ and $X_{\rm opt}$ is the optimal value of the
parameter, \ie the one belonging to the fittest combination of
parameters. We consider the mean value of $\Theta$ over the whole sample:
\begin{align}
	\bar{\Theta}=\frac{\sum^{N}_{\rm i} \Theta_{\rm i}}{N}\,,
\end{align}
where $N=N_{\rm pop}N_{\rm gen}$, $N_{\rm gen}$ being the total number of
generations and $N_{\rm pop}$ the number of population members in each
generation. We carry out this analysis for the five representative cases
\texttt{ALF2-M1225}, \texttt{APR4-M1350}, \texttt{GNH3-M1300},
\texttt{H4-M1275} and  \texttt{SLy-M1325}.

In all cases we find a value of $\bar{\Theta}<0.05$ for the disk mass
parameter $\mu_{\rm disk}$, the disk radius $\rho_{\rm disk}$ and the
length of the bar configuration $\lambda_{\rm bar}$. In other words these
parameters do not significantly evolve during the course of the
optimization process.

\pagebreak
To quantify the impact of the changes experienced by these parameters on
the value of the fitness function or the constraints accuracy, we vary
the optimal value of each of the four aforementioned parameters by its
relative variation along the generations, \ie we set $X_{\rm opt}\to
X'=X_{\rm opt}+X_{\rm opt}\bar{\Theta}$ and compute the corresponding
variation in the fitness function and constraints for the five
representative models listed above.

As a representative example, in the case of model \texttt{GNH3-M1300},
this procedure applied to $\mu_{\rm disk}$ and $\rho_{\rm disk}$ results
in a variation of $\rchitwo$ of a factor 100 and 50, respectively, while
the frequencies vary by less than 2\%.  Since $\lambda_{\rm bar}$ is the
parameter that influences $(dJ/dt)_{\rm bar}$ the most, varying it leads
to an increase of the value of $\rchitwo$ by a factor 140.  Analyzing the
impact of the remaining three parameters, we observe that even if the
damping constant $b$ has a mean $\bar{\Theta}$ of $\approx$ 8\%, this
leads to variations of $\rchitwo$ of a factor 0.5, and of $f_1$ and $f_2$
of less than 1\%.

At this point a second consideration comes into play. Examining
Eqs.~\eqref{eq:d} and \eqref{eq:omega}, it is apparent that quantities
such as \eg $M_{\rm disk}$ and $R_{\rm disk}$ are degenerate, appearing
in the equations only through their product. This means that although
small variations in $R_{\rm disk}$ can considerably vary $\rchitwo$, the
same holds true for $M_{\rm disk}$ (and $M_{\rm core}$). This
consideration does not apply, for instance, to $k$ and $b$ since they are
directly multiplied with $d$ or $\dot{d}$.

Summarizing, $\mu_{\rm disk}$ does not vary much during the optimization
process but its changes affect the fitness function greatly; $\rho_{\rm
disk}$ remains nearly constant and, although its variations may affect
$\rchitwo$, they can also be offset by $\mu_{\rm disk}$ (and $\mu_{\rm
core}$); the changes in $b$ do not modify significantly $\rchitwo$; and
varying $\lambda_{\rm bar}$ has a large impact on~$\rchitwo$.

We decided therefor to fix $\rho_{\rm disk}$ and $b$ to the constant
values of $3.5$ and $0.005$, respectively, which are
roughly the mean values of these parameters for the 5
representative models. The remaining parameters are determined by the GA,
for all EOSs and all mass configurations, with
$N_{\rm pop}=300$ and $N_{\rm gen}=100$. The GA results are
listed in Table~\ref{tab: GA_results}.

\bibliography{bibliography}

\begin{thebibliography}{73}%
\makeatletter
\providecommand \@ifxundefined [1]{%
 \@ifx{#1\undefined}
}%
\providecommand \@ifnum [1]{%
 \ifnum #1\expandafter \@firstoftwo
 \else \expandafter \@secondoftwo
 \fi
}%
\providecommand \@ifx [1]{%
 \ifx #1\expandafter \@firstoftwo
 \else \expandafter \@secondoftwo
 \fi
}%
\providecommand \natexlab [1]{#1}%
\providecommand \enquote  [1]{``#1''}%
\providecommand \bibnamefont  [1]{#1}%
\providecommand \bibfnamefont [1]{#1}%
\providecommand \citenamefont [1]{#1}%
\providecommand \href@noop [0]{\@secondoftwo}%
\providecommand \href [0]{\begingroup \@sanitize@url \@href}%
\providecommand \@href[1]{\@@startlink{#1}\@@href}%
\providecommand \@@href[1]{\endgroup#1\@@endlink}%
\providecommand \@sanitize@url [0]{\catcode `\\12\catcode `\$12\catcode
  `\&12\catcode `\#12\catcode `\^12\catcode `\_12\catcode `\%12\relax}%
\providecommand \@@startlink[1]{}%
\providecommand \@@endlink[0]{}%
\providecommand \url  [0]{\begingroup\@sanitize@url \@url }%
\providecommand \@url [1]{\endgroup\@href {#1}{\urlprefix }}%
\providecommand \urlprefix  [0]{URL }%
\providecommand \Eprint [0]{\href }%
\providecommand \doibase [0]{https://doi.org/}%
\providecommand \selectlanguage [0]{\@gobble}%
\providecommand \bibinfo  [0]{\@secondoftwo}%
\providecommand \bibfield  [0]{\@secondoftwo}%
\providecommand \translation [1]{[#1]}%
\providecommand \BibitemOpen [0]{}%
\providecommand \bibitemStop [0]{}%
\providecommand \bibitemNoStop [0]{.\EOS\space}%
\providecommand \EOS [0]{\spacefactor3000\relax}%
\providecommand \BibitemShut  [1]{\csname bibitem#1\endcsname}%
\let\auto@bib@innerbib\@empty
\bibitem [{\citenamefont {Giacconi}\ \emph {et~al.}(1962)\citenamefont
  {Giacconi}, \citenamefont {Gursky}, \citenamefont {Paolini},\ and\
  \citenamefont {Rossi}}]{PhysRevLett.9.439}%
  \BibitemOpen
  \bibfield  {author} {\bibinfo {author} {\bibfnamefont {R.}~\bibnamefont
  {Giacconi}}, \bibinfo {author} {\bibfnamefont {H.}~\bibnamefont {Gursky}},
  \bibinfo {author} {\bibfnamefont {F.~R.}\ \bibnamefont {Paolini}},\ and\
  \bibinfo {author} {\bibfnamefont {B.~B.}\ \bibnamefont {Rossi}},\ }\bibfield
  {title} {\bibinfo {title} {{Evidence for x Rays From Sources Outside the
  Solar System}},\ }\href {https://doi.org/10.1103/PhysRevLett.9.439}
  {\bibfield  {journal} {\bibinfo  {journal} {Phys. Rev. Lett.}\ }\textbf
  {\bibinfo {volume} {9}},\ \bibinfo {pages} {439} (\bibinfo {year}
  {1962})}\BibitemShut {NoStop}%
\bibitem [{\citenamefont {{Hewish}}\ \emph {et~al.}(1968)\citenamefont
  {{Hewish}}, \citenamefont {{Bell}}, \citenamefont {{Pilkington}},
  \citenamefont {{Scott}},\ and\ \citenamefont
  {{Collins}}}]{1968Natur.217..709H}%
  \BibitemOpen
  \bibfield  {author} {\bibinfo {author} {\bibfnamefont {A.}~\bibnamefont
  {{Hewish}}}, \bibinfo {author} {\bibfnamefont {S.~J.}\ \bibnamefont
  {{Bell}}}, \bibinfo {author} {\bibfnamefont {J.~D.~H.}\ \bibnamefont
  {{Pilkington}}}, \bibinfo {author} {\bibfnamefont {P.~F.}\ \bibnamefont
  {{Scott}}},\ and\ \bibinfo {author} {\bibfnamefont {R.~A.}\ \bibnamefont
  {{Collins}}},\ }\bibfield  {title} {\bibinfo {title} {{Observation of a
  Rapidly Pulsating Radio Source}},\ }\href {https://doi.org/10.1038/217709a0}
  {\bibfield  {journal} {\bibinfo  {journal} {Nature}\ }\textbf {\bibinfo
  {volume} {217}},\ \bibinfo {pages} {709} (\bibinfo {year}
  {1968})}\BibitemShut {NoStop}%
\bibitem [{\citenamefont {Yakovlev}\ \emph {et~al.}(2013)\citenamefont
  {Yakovlev}, \citenamefont {Haensel}, \citenamefont {Baym},\ and\
  \citenamefont {Pethick}}]{Yakovlev_2013}%
  \BibitemOpen
  \bibfield  {author} {\bibinfo {author} {\bibfnamefont {D.~G.}\ \bibnamefont
  {Yakovlev}}, \bibinfo {author} {\bibfnamefont {P.}~\bibnamefont {Haensel}},
  \bibinfo {author} {\bibfnamefont {G.}~\bibnamefont {Baym}},\ and\ \bibinfo
  {author} {\bibfnamefont {C.}~\bibnamefont {Pethick}},\ }\bibfield  {title}
  {\bibinfo {title} {{Lev Landau and the concept of neutron stars}},\ }\href
  {https://doi.org/10.3367/ufne.0183.201303f.0307} {\bibfield  {journal}
  {\bibinfo  {journal} {Physics-Uspekhi}\ }\textbf {\bibinfo {volume} {56}},\
  \bibinfo {pages} {289} (\bibinfo {year} {2013})}\BibitemShut {NoStop}%
\bibitem [{\citenamefont {Baym}(1982)}]{schofield1982neutron}%
  \BibitemOpen
  \bibfield  {author} {\bibinfo {author} {\bibfnamefont {G.}~\bibnamefont
  {Baym}},\ }\bibfield  {title} {\bibinfo {title} {Neutron stars: the first
  fifty years},\ }in\ \href@noop {} {\emph {\bibinfo {booktitle} {The Neutron
  and its Applications}}}\ (\bibinfo {year} {1982})\BibitemShut {NoStop}%
\bibitem [{\citenamefont {{Hulse}}\ and\ \citenamefont
  {{Taylor}}(1975)}]{1975ApJ...195L..51H}%
  \BibitemOpen
  \bibfield  {author} {\bibinfo {author} {\bibfnamefont {R.~A.}\ \bibnamefont
  {{Hulse}}}\ and\ \bibinfo {author} {\bibfnamefont {J.~H.}\ \bibnamefont
  {{Taylor}}},\ }\bibfield  {title} {\bibinfo {title} {Discovery of a pulsar in
  a binary system},\ }\href {https://doi.org/10.1086/181708} {\bibfield
  {journal} {\bibinfo  {journal} {Astrophysical Journal}\ }\textbf {\bibinfo
  {volume} {195}},\ \bibinfo {pages} {L51} (\bibinfo {year}
  {1975})}\BibitemShut {NoStop}%
\bibitem [{\citenamefont {{Eichler}}\ \emph {et~al.}(1989)\citenamefont
  {{Eichler}}, \citenamefont {{Livio}}, \citenamefont {{Piran}},\ and\
  \citenamefont {{Schramm}}}]{Eichler89}%
  \BibitemOpen
  \bibfield  {author} {\bibinfo {author} {\bibfnamefont {D.}~\bibnamefont
  {{Eichler}}}, \bibinfo {author} {\bibfnamefont {M.}~\bibnamefont {{Livio}}},
  \bibinfo {author} {\bibfnamefont {T.}~\bibnamefont {{Piran}}},\ and\ \bibinfo
  {author} {\bibfnamefont {D.~N.}\ \bibnamefont {{Schramm}}},\ }\bibfield
  {title} {\bibinfo {title} {{Nucleosynthesis, neutrino bursts and gamma-rays
  from coalescing neutron stars}},\ }\href {https://doi.org/10.1038/340126a0}
  {\bibfield  {journal} {\bibinfo  {journal} {Nature}\ }\textbf {\bibinfo
  {volume} {340}},\ \bibinfo {pages} {126} (\bibinfo {year}
  {1989})}\BibitemShut {NoStop}%
\bibitem [{\citenamefont {{Narayan}}\ \emph {et~al.}(1992)\citenamefont
  {{Narayan}}, \citenamefont {{Paczynski}},\ and\ \citenamefont
  {{Piran}}}]{Narayan92}%
  \BibitemOpen
  \bibfield  {author} {\bibinfo {author} {\bibfnamefont {R.}~\bibnamefont
  {{Narayan}}}, \bibinfo {author} {\bibfnamefont {B.}~\bibnamefont
  {{Paczynski}}},\ and\ \bibinfo {author} {\bibfnamefont {T.}~\bibnamefont
  {{Piran}}},\ }\bibfield  {title} {\bibinfo {title} {{Gamma-ray bursts as the
  death throes of massive binary stars}},\ }\href
  {https://doi.org/10.1086/186493} {\bibfield  {journal} {\bibinfo  {journal}
  {Astrophys. J. Lett.}\ }\textbf {\bibinfo {volume} {395}},\ \bibinfo {pages}
  {L83} (\bibinfo {year} {1992})},\ \Eprint
  {https://arxiv.org/abs/astro-ph/9204001} {astro-ph/9204001} \BibitemShut
  {NoStop}%
\bibitem [{\citenamefont {{LIGO collaboration}}()}]{LIGO_history}%
  \BibitemOpen
  \bibfield  {author} {\bibinfo {author} {\bibnamefont {{LIGO
  collaboration}}},\ }\href@noop {} {\bibinfo {title} {{A Brief History of
  LIGO}}},\ \bibinfo {howpublished}
  {\url{https://www.ligo.caltech.edu/system/media_files/binaries/313/original/LIGOHistory.pdf}},\
  \bibinfo {note} {accessed: 2019-10-28}\BibitemShut {NoStop}%
\bibitem [{\citenamefont {{Gamma-Ray Astrophysics Team,
  NSSTC}}()}]{Fermi_GMB_history}%
  \BibitemOpen
  \bibfield  {author} {\bibinfo {author} {\bibnamefont {{Gamma-Ray Astrophysics
  Team, NSSTC}}},\ }\href@noop {} {\bibinfo {title} {{Fermi GBM}}},\ \bibinfo
  {howpublished} {\url{https://gammaray.msfc.nasa.gov/gbm/}},\ \bibinfo {note}
  {accessed: 2019-10-28}\BibitemShut {NoStop}%
\bibitem [{\citenamefont {{Abbott}}\ \emph
  {et~al.}(2017{\natexlab{a}})\citenamefont {{Abbott}} \emph
  {et~al.}}]{abbott2017b_NS}%
  \BibitemOpen
  \bibfield  {author} {\bibinfo {author} {\bibfnamefont {B.~P.}\ \bibnamefont
  {{Abbott}}} \emph {et~al.} (\bibinfo {collaboration} {LIGO Scientific
  Collaboration and Virgo Collaboration}),\ }\bibfield  {title} {\bibinfo
  {title} {{Multi-messenger Observations of a Binary Neutron Star Merger}},\
  }\href {http://stacks.iop.org/2041-8205/848/i=2/a=L12} {\bibfield  {journal}
  {\bibinfo  {journal} {Astrophys. J. Lett.}\ }\textbf {\bibinfo {volume}
  {848}},\ \bibinfo {pages} {L12} (\bibinfo {year}
  {2017}{\natexlab{a}})}\BibitemShut {NoStop}%
\bibitem [{\citenamefont {{Abbott}}\ \emph
  {et~al.}(2017{\natexlab{b}})\citenamefont {{Abbott}} \emph
  {et~al.}}]{abbott2017d_NS}%
  \BibitemOpen
  \bibfield  {author} {\bibinfo {author} {\bibfnamefont {B.~P.}\ \bibnamefont
  {{Abbott}}} \emph {et~al.} (\bibinfo {collaboration} {LIGO Scientific
  Collaboration and Virgo Collaboration}),\ }\bibfield  {title} {\bibinfo
  {title} {{Gravitational Waves and Gamma-Rays from a Binary Neutron Star
  Merger: GW170817 and GRB 170817A}},\ }\href
  {http://stacks.iop.org/2041-8205/848/i=2/a=L13} {\bibfield  {journal}
  {\bibinfo  {journal} {Astrophys. J. Lett.}\ }\textbf {\bibinfo {volume}
  {848}},\ \bibinfo {pages} {L13} (\bibinfo {year} {2017}{\natexlab{b}})},\
  \Eprint {https://arxiv.org/abs/1710.05834} {arXiv:1710.05834 [astro-ph.HE]}
  \BibitemShut {NoStop}%
\bibitem [{\citenamefont {{Abbott}}\ \emph
  {et~al.}(2017{\natexlab{c}})\citenamefont {{Abbott}} \emph
  {et~al.}}]{abbott2017a_NS}%
  \BibitemOpen
  \bibfield  {author} {\bibinfo {author} {\bibfnamefont {B.~P.}\ \bibnamefont
  {{Abbott}}} \emph {et~al.} (\bibinfo {collaboration} {LIGO Scientific
  Collaboration and Virgo Collaboration}),\ }\bibfield  {title} {\bibinfo
  {title} {{GW170817: Observation of Gravitational Waves from a Binary Neutron
  Star Inspiral}},\ }\href {https://doi.org/10.1103/PhysRevLett.119.161101}
  {\bibfield  {journal} {\bibinfo  {journal} {Phys. Rev. Lett.}\ }\textbf
  {\bibinfo {volume} {119}},\ \bibinfo {pages} {161101} (\bibinfo {year}
  {2017}{\natexlab{c}})}\BibitemShut {NoStop}%
\bibitem [{\citenamefont {{Goldstein}}\ \emph {et~al.}(2017)\citenamefont
  {{Goldstein}} \emph {et~al.}}]{Goldstein2017}%
  \BibitemOpen
  \bibfield  {author} {\bibinfo {author} {\bibfnamefont {A.}~\bibnamefont
  {{Goldstein}}} \emph {et~al.},\ }\bibfield  {title} {\bibinfo {title} {{An
  Ordinary Short Gamma-Ray Burst with Extraordinary Implications: Fermi-GBM
  Detection of GRB 170817A}},\ }\href
  {https://doi.org/10.3847/2041-8213/aa8f41} {\bibfield  {journal} {\bibinfo
  {journal} {Astrophys. J. Letters}\ }\textbf {\bibinfo {volume} {848}},\
  \bibinfo {eid} {L14} (\bibinfo {year} {2017})},\ \Eprint
  {https://arxiv.org/abs/1710.05446} {arXiv:1710.05446 [astro-ph.HE]}
  \BibitemShut {NoStop}%
\bibitem [{\citenamefont {{Savchenko}}\ \emph {et~al.}(2017)\citenamefont
  {{Savchenko}} \emph {et~al.}}]{Savchenko2017}%
  \BibitemOpen
  \bibfield  {author} {\bibinfo {author} {\bibfnamefont {V.}~\bibnamefont
  {{Savchenko}}} \emph {et~al.},\ }\bibfield  {title} {\bibinfo {title}
  {{INTEGRAL Detection of the First Prompt Gamma-Ray Signal Coincident with the
  Gravitational-wave Event GW170817}},\ }\href
  {https://doi.org/10.3847/2041-8213/aa8f94} {\bibfield  {journal} {\bibinfo
  {journal} {Astrophys. J. Letters}\ }\textbf {\bibinfo {volume} {848}},\
  \bibinfo {eid} {L15} (\bibinfo {year} {2017})},\ \Eprint
  {https://arxiv.org/abs/1710.05449} {arXiv:1710.05449 [astro-ph.HE]}
  \BibitemShut {NoStop}%
\bibitem [{\citenamefont {{Coulter}}\ and\ \citenamefont
  {et~al.}(2017)}]{Coulter2017}%
  \BibitemOpen
  \bibfield  {author} {\bibinfo {author} {\bibfnamefont {D.~A.}\ \bibnamefont
  {{Coulter}}}\ and\ \bibinfo {author} {\bibnamefont {et~al.}},\ }\bibfield
  {title} {\bibinfo {title} {{Swope Supernova Survey 2017a (SSS17a), the
  optical counterpart to a gravitational wave source}},\ }\href
  {https://doi.org/10.1126/science.aap9811} {\bibfield  {journal} {\bibinfo
  {journal} {Science}\ }\textbf {\bibinfo {volume} {358}},\ \bibinfo {pages}
  {1556} (\bibinfo {year} {2017})},\ \Eprint {https://arxiv.org/abs/1710.05452}
  {arXiv:1710.05452 [astro-ph.HE]} \BibitemShut {NoStop}%
\bibitem [{\citenamefont {{Abbott}}\ \emph
  {et~al.}(2017{\natexlab{d}})\citenamefont {{Abbott}} \emph
  {et~al.}}]{abbott2017c_NS}%
  \BibitemOpen
  \bibfield  {author} {\bibinfo {author} {\bibfnamefont {B.~P.}\ \bibnamefont
  {{Abbott}}} \emph {et~al.} (\bibinfo {collaboration} {LIGO Scientific
  Collaboration and Virgo Collaboration}),\ }\bibfield  {title} {\bibinfo
  {title} {{Estimating the Contribution of Dynamical Ejecta in the Kilonova
  Associated with GW170817}},\ }\href
  {https://doi.org/10.3847/2041-8213/aa9478} {\bibfield  {journal} {\bibinfo
  {journal} {Astrophys. J. Lett.}\ }\textbf {\bibinfo {volume} {850}},\
  \bibinfo {eid} {L39} (\bibinfo {year} {2017}{\natexlab{d}})},\ \Eprint
  {https://arxiv.org/abs/1710.05836} {arXiv:1710.05836 [astro-ph.HE]}
  \BibitemShut {NoStop}%
\bibitem [{\citenamefont {{Rasio}}\ and\ \citenamefont
  {{Shapiro}}(1992)}]{Rasio92}%
  \BibitemOpen
  \bibfield  {author} {\bibinfo {author} {\bibfnamefont {F.~A.}\ \bibnamefont
  {{Rasio}}}\ and\ \bibinfo {author} {\bibfnamefont {S.~L.}\ \bibnamefont
  {{Shapiro}}},\ }\bibfield  {title} {\bibinfo {title} {{Hydrodynamical
  evolution of coalescing binary neutron stars}},\ }\href
  {https://doi.org/10.1086/172055} {\bibfield  {journal} {\bibinfo  {journal}
  {Astrophys. J.}\ }\textbf {\bibinfo {volume} {401}},\ \bibinfo {pages} {226}
  (\bibinfo {year} {1992})}\BibitemShut {NoStop}%
\bibitem [{\citenamefont {{Ruffert}}\ and\ \citenamefont
  {{Melia}}(1994)}]{Ruffert1994}%
  \BibitemOpen
  \bibfield  {author} {\bibinfo {author} {\bibfnamefont {M.}~\bibnamefont
  {{Ruffert}}}\ and\ \bibinfo {author} {\bibfnamefont {F.}~\bibnamefont
  {{Melia}}},\ }\bibfield  {title} {\bibinfo {title} {{Hydrodynamical 3D
  Bondi-Hoyle accretion onto the Galactic Center blackhole candidate SGR A*}},\
  }\href@noop {} {\bibfield  {journal} {\bibinfo  {journal} {Astron.
  Astrophys.}\ }\textbf {\bibinfo {volume} {288}},\ \bibinfo {pages} {L29}
  (\bibinfo {year} {1994})}\BibitemShut {NoStop}%
\bibitem [{\citenamefont {Shibata}\ and\ \citenamefont
  {Uryu}(2000)}]{Shibata:1999wm}%
  \BibitemOpen
  \bibfield  {author} {\bibinfo {author} {\bibfnamefont {M.}~\bibnamefont
  {Shibata}}\ and\ \bibinfo {author} {\bibfnamefont {K.}~\bibnamefont {Uryu}},\
  }\bibfield  {title} {\bibinfo {title} {{Simulation of merging binary neutron
  stars in full general relativity: $\Gamma = 2$ case}},\ }\href
  {https://doi.org/10.1103/PhysRevD.61.064001} {\bibfield  {journal} {\bibinfo
  {journal} {Phys. Rev.}\ }\textbf {\bibinfo {volume} {D61}},\ \bibinfo {pages}
  {064001} (\bibinfo {year} {2000})},\ \Eprint
  {https://arxiv.org/abs/gr-qc/9911058} {arXiv:gr-qc/9911058 [gr-qc]}
  \BibitemShut {NoStop}%
\bibitem [{\citenamefont {Shibata}\ and\ \citenamefont
  {Ury\=u}(2001)}]{Shibata01a}%
  \BibitemOpen
  \bibfield  {author} {\bibinfo {author} {\bibfnamefont {M.}~\bibnamefont
  {Shibata}}\ and\ \bibinfo {author} {\bibfnamefont {K.}~\bibnamefont
  {Ury\=u}},\ }\bibfield  {title} {\bibinfo {title} {{Computation of
  gravitational waves from inspiraling binary neutron stars in quasiequilibrium
  circular orbits: Formulation and calibration}},\ }\href@noop {} {\bibfield
  {journal} {\bibinfo  {journal} {Phys. Rev. D}\ }\textbf {\bibinfo {volume}
  {64}},\ \bibinfo {pages} {104017} (\bibinfo {year} {2001})}\BibitemShut
  {NoStop}%
\bibitem [{\citenamefont {Shibata}\ and\ \citenamefont
  {Uryu}(2002)}]{Shibata:2002jb}%
  \BibitemOpen
  \bibfield  {author} {\bibinfo {author} {\bibfnamefont {M.}~\bibnamefont
  {Shibata}}\ and\ \bibinfo {author} {\bibfnamefont {K.}~\bibnamefont {Uryu}},\
  }\bibfield  {title} {\bibinfo {title} {{Gravitational waves from the merger
  of binary neutron stars in a fully general relativistic simulation}},\ }\href
  {https://doi.org/10.1143/PTP.107.265} {\bibfield  {journal} {\bibinfo
  {journal} {Prog. Theor. Phys.}\ }\textbf {\bibinfo {volume} {107}},\ \bibinfo
  {pages} {265} (\bibinfo {year} {2002})},\ \Eprint
  {https://arxiv.org/abs/gr-qc/0203037} {arXiv:gr-qc/0203037 [gr-qc]}
  \BibitemShut {NoStop}%
\bibitem [{\citenamefont {Font}\ \emph {et~al.}(2002)\citenamefont {Font},
  \citenamefont {Goodale}, \citenamefont {Iyer}, \citenamefont {Miller},
  \citenamefont {Rezzolla}, \citenamefont {Seidel}, \citenamefont
  {Stergioulas}, \citenamefont {Suen},\ and\ \citenamefont
  {Tobias}}]{PhysRevD.65.084024}%
  \BibitemOpen
  \bibfield  {author} {\bibinfo {author} {\bibfnamefont {J.~A.}\ \bibnamefont
  {Font}}, \bibinfo {author} {\bibfnamefont {T.}~\bibnamefont {Goodale}},
  \bibinfo {author} {\bibfnamefont {S.}~\bibnamefont {Iyer}}, \bibinfo {author}
  {\bibfnamefont {M.}~\bibnamefont {Miller}}, \bibinfo {author} {\bibfnamefont
  {L.}~\bibnamefont {Rezzolla}}, \bibinfo {author} {\bibfnamefont
  {E.}~\bibnamefont {Seidel}}, \bibinfo {author} {\bibfnamefont
  {N.}~\bibnamefont {Stergioulas}}, \bibinfo {author} {\bibfnamefont {W.-M.}\
  \bibnamefont {Suen}},\ and\ \bibinfo {author} {\bibfnamefont
  {M.}~\bibnamefont {Tobias}},\ }\bibfield  {title} {\bibinfo {title}
  {Three-dimensional numerical general relativistic hydrodynamics. ii.
  long-term dynamics of single relativistic stars},\ }\href
  {https://doi.org/10.1103/PhysRevD.65.084024} {\bibfield  {journal} {\bibinfo
  {journal} {Phys. Rev. D}\ }\textbf {\bibinfo {volume} {65}},\ \bibinfo
  {pages} {084024} (\bibinfo {year} {2002})}\BibitemShut {NoStop}%
\bibitem [{\citenamefont {Shibata}\ and\ \citenamefont
  {Taniguchi}(2006)}]{Shibata:2006nm}%
  \BibitemOpen
  \bibfield  {author} {\bibinfo {author} {\bibfnamefont {M.}~\bibnamefont
  {Shibata}}\ and\ \bibinfo {author} {\bibfnamefont {K.}~\bibnamefont
  {Taniguchi}},\ }\bibfield  {title} {\bibinfo {title} {{Merger of binary
  neutron stars to a black hole: disk mass, short gamma-ray bursts, and
  quasinormal mode ringing}},\ }\href
  {https://doi.org/10.1103/PhysRevD.73.064027} {\bibfield  {journal} {\bibinfo
  {journal} {Phys. Rev.}\ }\textbf {\bibinfo {volume} {D73}},\ \bibinfo {pages}
  {064027} (\bibinfo {year} {2006})},\ \Eprint
  {https://arxiv.org/abs/astro-ph/0603145} {arXiv:astro-ph/0603145 [astro-ph]}
  \BibitemShut {NoStop}%
\bibitem [{\citenamefont {Anderson}\ \emph {et~al.}(2008)\citenamefont
  {Anderson}, \citenamefont {Hirschmann}, \citenamefont {Lehner}, \citenamefont
  {Liebling}, \citenamefont {Motl}, \citenamefont {Neilsen}, \citenamefont
  {Palenzuela},\ and\ \citenamefont {Tohline}}]{Anderson:2007kz}%
  \BibitemOpen
  \bibfield  {author} {\bibinfo {author} {\bibfnamefont {M.}~\bibnamefont
  {Anderson}}, \bibinfo {author} {\bibfnamefont {E.~W.}\ \bibnamefont
  {Hirschmann}}, \bibinfo {author} {\bibfnamefont {L.}~\bibnamefont {Lehner}},
  \bibinfo {author} {\bibfnamefont {S.~L.}\ \bibnamefont {Liebling}}, \bibinfo
  {author} {\bibfnamefont {P.~M.}\ \bibnamefont {Motl}}, \bibinfo {author}
  {\bibfnamefont {D.}~\bibnamefont {Neilsen}}, \bibinfo {author} {\bibfnamefont
  {C.}~\bibnamefont {Palenzuela}},\ and\ \bibinfo {author} {\bibfnamefont
  {J.~E.}\ \bibnamefont {Tohline}},\ }\bibfield  {title} {\bibinfo {title}
  {{Simulating binary neutron stars: Dynamics and gravitational waves}},\
  }\href {https://doi.org/10.1103/PhysRevD.77.024006} {\bibfield  {journal}
  {\bibinfo  {journal} {Phys. Rev.}\ }\textbf {\bibinfo {volume} {D77}},\
  \bibinfo {pages} {024006} (\bibinfo {year} {2008})},\ \Eprint
  {https://arxiv.org/abs/0708.2720} {arXiv:0708.2720 [gr-qc]} \BibitemShut
  {NoStop}%
\bibitem [{\citenamefont {Yamamoto}\ \emph {et~al.}(2008)\citenamefont
  {Yamamoto}, \citenamefont {Shibata},\ and\ \citenamefont
  {Taniguchi}}]{Yamamoto:2008js}%
  \BibitemOpen
  \bibfield  {author} {\bibinfo {author} {\bibfnamefont {T.}~\bibnamefont
  {Yamamoto}}, \bibinfo {author} {\bibfnamefont {M.}~\bibnamefont {Shibata}},\
  and\ \bibinfo {author} {\bibfnamefont {K.}~\bibnamefont {Taniguchi}},\
  }\bibfield  {title} {\bibinfo {title} {{Simulating coalescing compact
  binaries by a new code SACRA}},\ }\href
  {https://doi.org/10.1103/PhysRevD.78.064054} {\bibfield  {journal} {\bibinfo
  {journal} {Phys. Rev.}\ }\textbf {\bibinfo {volume} {D78}},\ \bibinfo {pages}
  {064054} (\bibinfo {year} {2008})},\ \Eprint
  {https://arxiv.org/abs/0806.4007} {arXiv:0806.4007 [gr-qc]} \BibitemShut
  {NoStop}%
\bibitem [{\citenamefont {{Baiotti}}\ \emph {et~al.}(2008)\citenamefont
  {{Baiotti}}, \citenamefont {{Giacomazzo}},\ and\ \citenamefont
  {{Rezzolla}}}]{Baiotti08}%
  \BibitemOpen
  \bibfield  {author} {\bibinfo {author} {\bibfnamefont {L.}~\bibnamefont
  {{Baiotti}}}, \bibinfo {author} {\bibfnamefont {B.}~\bibnamefont
  {{Giacomazzo}}},\ and\ \bibinfo {author} {\bibfnamefont {L.}~\bibnamefont
  {{Rezzolla}}},\ }\bibfield  {title} {\bibinfo {title} {{Accurate evolutions
  of inspiralling neutron-star binaries: Prompt and delayed collapse to a black
  hole}},\ }\href {https://doi.org/10.1103/PhysRevD.78.084033} {\bibfield
  {journal} {\bibinfo  {journal} {Phys. Rev. D}\ }\textbf {\bibinfo {volume}
  {78}},\ \bibinfo {pages} {084033} (\bibinfo {year} {2008})},\ \Eprint
  {https://arxiv.org/abs/0804.0594} {arXiv:0804.0594 [gr-qc]} \BibitemShut
  {NoStop}%
\bibitem [{\citenamefont {{Rezzolla}}\ \emph {et~al.}(2010)\citenamefont
  {{Rezzolla}}, \citenamefont {{Baiotti}}, \citenamefont {{Giacomazzo}},
  \citenamefont {{Link}},\ and\ \citenamefont {{Font}}}]{Rezzolla:2010}%
  \BibitemOpen
  \bibfield  {author} {\bibinfo {author} {\bibfnamefont {L.}~\bibnamefont
  {{Rezzolla}}}, \bibinfo {author} {\bibfnamefont {L.}~\bibnamefont
  {{Baiotti}}}, \bibinfo {author} {\bibfnamefont {B.}~\bibnamefont
  {{Giacomazzo}}}, \bibinfo {author} {\bibfnamefont {D.}~\bibnamefont
  {{Link}}},\ and\ \bibinfo {author} {\bibfnamefont {J.~A.}\ \bibnamefont
  {{Font}}},\ }\bibfield  {title} {\bibinfo {title} {{Accurate evolutions of
  unequal-mass neutron-star binaries: properties of the torus and short GRB
  engines}},\ }\href {https://doi.org/10.1088/0264-9381/27/11/114105}
  {\bibfield  {journal} {\bibinfo  {journal} {Class. Quantum Grav.}\ }\textbf
  {\bibinfo {volume} {27}},\ \bibinfo {pages} {114105} (\bibinfo {year}
  {2010})},\ \Eprint {https://arxiv.org/abs/1001.3074} {arXiv:1001.3074
  [gr-qc]} \BibitemShut {NoStop}%
\bibitem [{\citenamefont {{Anderson}}\ \emph {et~al.}(2008)\citenamefont
  {{Anderson}}, \citenamefont {{Hirschmann}}, \citenamefont {{Lehner}},
  \citenamefont {{Liebling}}, \citenamefont {{Motl}}, \citenamefont
  {{Neilsen}}, \citenamefont {{Palenzuela}},\ and\ \citenamefont
  {{Tohline}}}]{Anderson2008}%
  \BibitemOpen
  \bibfield  {author} {\bibinfo {author} {\bibfnamefont {M.}~\bibnamefont
  {{Anderson}}}, \bibinfo {author} {\bibfnamefont {E.~W.}\ \bibnamefont
  {{Hirschmann}}}, \bibinfo {author} {\bibfnamefont {L.}~\bibnamefont
  {{Lehner}}}, \bibinfo {author} {\bibfnamefont {S.~L.}\ \bibnamefont
  {{Liebling}}}, \bibinfo {author} {\bibfnamefont {P.~M.}\ \bibnamefont
  {{Motl}}}, \bibinfo {author} {\bibfnamefont {D.}~\bibnamefont {{Neilsen}}},
  \bibinfo {author} {\bibfnamefont {C.}~\bibnamefont {{Palenzuela}}},\ and\
  \bibinfo {author} {\bibfnamefont {J.~E.}\ \bibnamefont {{Tohline}}},\
  }\bibfield  {title} {\bibinfo {title} {Magnetized neutron-star mergers and
  gravitational-wave signals},\ }\href
  {https://doi.org/10.1103/PhysRevLett.100.191101} {\bibfield  {journal}
  {\bibinfo  {journal} {Phys. Rev. Lett.}\ }\textbf {\bibinfo {volume} {100}},\
  \bibinfo {eid} {191101} (\bibinfo {year} {2008})},\ \Eprint
  {https://arxiv.org/abs/0801.4387} {arXiv:0801.4387 [gr-qc]} \BibitemShut
  {NoStop}%
\bibitem [{\citenamefont {{Giacomazzo}}\ \emph {et~al.}(2011)\citenamefont
  {{Giacomazzo}}, \citenamefont {{Rezzolla}},\ and\ \citenamefont
  {{Baiotti}}}]{Giacomazzo2011b}%
  \BibitemOpen
  \bibfield  {author} {\bibinfo {author} {\bibfnamefont {B.}~\bibnamefont
  {{Giacomazzo}}}, \bibinfo {author} {\bibfnamefont {L.}~\bibnamefont
  {{Rezzolla}}},\ and\ \bibinfo {author} {\bibfnamefont {L.}~\bibnamefont
  {{Baiotti}}},\ }\bibfield  {title} {\bibinfo {title} {Accurate evolutions of
  inspiralling and magnetized neutron stars: Equal-mass binaries},\ }\href
  {https://doi.org/10.1103/PhysRevD.83.044014} {\bibfield  {journal} {\bibinfo
  {journal} {Phys. Rev. D}\ }\textbf {\bibinfo {volume} {83}},\ \bibinfo {eid}
  {044014} (\bibinfo {year} {2011})},\ \Eprint
  {https://arxiv.org/abs/1009.2468} {arXiv:1009.2468 [gr-qc]} \BibitemShut
  {NoStop}%
\bibitem [{\citenamefont {{Ciolfi}}\ \emph {et~al.}(2017)\citenamefont
  {{Ciolfi}}, \citenamefont {{Kastaun}}, \citenamefont {{Giacomazzo}},
  \citenamefont {{Endrizzi}}, \citenamefont {{Siegel}},\ and\ \citenamefont
  {{Perna}}}]{Ciolfi2017}%
  \BibitemOpen
  \bibfield  {author} {\bibinfo {author} {\bibfnamefont {R.}~\bibnamefont
  {{Ciolfi}}}, \bibinfo {author} {\bibfnamefont {W.}~\bibnamefont {{Kastaun}}},
  \bibinfo {author} {\bibfnamefont {B.}~\bibnamefont {{Giacomazzo}}}, \bibinfo
  {author} {\bibfnamefont {A.}~\bibnamefont {{Endrizzi}}}, \bibinfo {author}
  {\bibfnamefont {D.~M.}\ \bibnamefont {{Siegel}}},\ and\ \bibinfo {author}
  {\bibfnamefont {R.}~\bibnamefont {{Perna}}},\ }\bibfield  {title} {\bibinfo
  {title} {{General relativistic magnetohydrodynamic simulations of binary
  neutron star mergers forming a long-lived neutron star}},\ }\href
  {https://doi.org/10.1103/PhysRevD.95.063016} {\bibfield  {journal} {\bibinfo
  {journal} {Phys. Rev. D}\ }\textbf {\bibinfo {volume} {95}},\ \bibinfo {eid}
  {063016} (\bibinfo {year} {2017})},\ \Eprint
  {https://arxiv.org/abs/1701.08738} {arXiv:1701.08738 [astro-ph.HE]}
  \BibitemShut {NoStop}%
\bibitem [{\citenamefont {{Kaplan}}\ \emph {et~al.}(2014)\citenamefont
  {{Kaplan}}, \citenamefont {{Ott}}, \citenamefont {{O'Connor}}, \citenamefont
  {{Kiuchi}}, \citenamefont {{Roberts}},\ and\ \citenamefont
  {{Duez}}}]{Kaplan2013}%
  \BibitemOpen
  \bibfield  {author} {\bibinfo {author} {\bibfnamefont {J.~D.}\ \bibnamefont
  {{Kaplan}}}, \bibinfo {author} {\bibfnamefont {C.~D.}\ \bibnamefont {{Ott}}},
  \bibinfo {author} {\bibfnamefont {E.~P.}\ \bibnamefont {{O'Connor}}},
  \bibinfo {author} {\bibfnamefont {K.}~\bibnamefont {{Kiuchi}}}, \bibinfo
  {author} {\bibfnamefont {L.}~\bibnamefont {{Roberts}}},\ and\ \bibinfo
  {author} {\bibfnamefont {M.}~\bibnamefont {{Duez}}},\ }\bibfield  {title}
  {\bibinfo {title} {The influence of thermal pressure on equilibrium models of
  hypermassive neutron star merger remnants},\ }\href
  {https://doi.org/10.1088/0004-637X/790/1/19} {\bibfield  {journal} {\bibinfo
  {journal} {Astrophys. J.}\ }\textbf {\bibinfo {volume} {790}},\ \bibinfo
  {eid} {19} (\bibinfo {year} {2014})},\ \Eprint
  {https://arxiv.org/abs/1306.4034} {arXiv:1306.4034 [astro-ph.HE]}
  \BibitemShut {NoStop}%
\bibitem [{\citenamefont {Endrizzi}\ \emph {et~al.}(2020)\citenamefont
  {Endrizzi}, \citenamefont {Perego}, \citenamefont {Fabbri}, \citenamefont
  {Branca}, \citenamefont {Radice}, \citenamefont {Bernuzzi}, \citenamefont
  {Giacomazzo}, \citenamefont {Pederiva},\ and\ \citenamefont
  {Lovato}}]{Endrizzi:2019trv}%
  \BibitemOpen
  \bibfield  {author} {\bibinfo {author} {\bibfnamefont {A.}~\bibnamefont
  {Endrizzi}}, \bibinfo {author} {\bibfnamefont {A.}~\bibnamefont {Perego}},
  \bibinfo {author} {\bibfnamefont {F.~M.}\ \bibnamefont {Fabbri}}, \bibinfo
  {author} {\bibfnamefont {L.}~\bibnamefont {Branca}}, \bibinfo {author}
  {\bibfnamefont {D.}~\bibnamefont {Radice}}, \bibinfo {author} {\bibfnamefont
  {S.}~\bibnamefont {Bernuzzi}}, \bibinfo {author} {\bibfnamefont
  {B.}~\bibnamefont {Giacomazzo}}, \bibinfo {author} {\bibfnamefont
  {F.}~\bibnamefont {Pederiva}},\ and\ \bibinfo {author} {\bibfnamefont
  {A.}~\bibnamefont {Lovato}},\ }\bibfield  {title} {\bibinfo {title}
  {{Thermodynamics conditions of matter in the neutrino decoupling region
  during neutron star mergers}},\ }\href
  {https://doi.org/10.1140/epja/s10050-019-00018-6} {\bibfield  {journal}
  {\bibinfo  {journal} {Eur. Phys. J. A}\ }\textbf {\bibinfo {volume} {56}},\
  \bibinfo {pages} {15} (\bibinfo {year} {2020})},\ \Eprint
  {https://arxiv.org/abs/1908.04952} {arXiv:1908.04952 [astro-ph.HE]}
  \BibitemShut {NoStop}%
\bibitem [{\citenamefont {{Kastaun}}\ and\ \citenamefont
  {{Galeazzi}}(2015)}]{Kastaun2014}%
  \BibitemOpen
  \bibfield  {author} {\bibinfo {author} {\bibfnamefont {W.}~\bibnamefont
  {{Kastaun}}}\ and\ \bibinfo {author} {\bibfnamefont {F.}~\bibnamefont
  {{Galeazzi}}},\ }\bibfield  {title} {\bibinfo {title} {Properties of
  hypermassive neutron stars formed in mergers of spinning binaries},\ }\href
  {https://doi.org/10.1103/PhysRevD.91.064027} {\bibfield  {journal} {\bibinfo
  {journal} {Phys. Rev. D}\ }\textbf {\bibinfo {volume} {91}},\ \bibinfo {eid}
  {064027} (\bibinfo {year} {2015})},\ \Eprint
  {https://arxiv.org/abs/1411.7975} {arXiv:1411.7975 [gr-qc]} \BibitemShut
  {NoStop}%
\bibitem [{\citenamefont {{Bernuzzi}}\ \emph {et~al.}(2016)\citenamefont
  {{Bernuzzi}}, \citenamefont {{Radice}}, \citenamefont {{Ott}}, \citenamefont
  {{Roberts}}, \citenamefont {{Moesta}},\ and\ \citenamefont
  {{Galeazzi}}}]{Bernuzzi2015b}%
  \BibitemOpen
  \bibfield  {author} {\bibinfo {author} {\bibfnamefont {S.}~\bibnamefont
  {{Bernuzzi}}}, \bibinfo {author} {\bibfnamefont {D.}~\bibnamefont
  {{Radice}}}, \bibinfo {author} {\bibfnamefont {C.~D.}\ \bibnamefont {{Ott}}},
  \bibinfo {author} {\bibfnamefont {L.~F.}\ \bibnamefont {{Roberts}}}, \bibinfo
  {author} {\bibfnamefont {P.}~\bibnamefont {{Moesta}}},\ and\ \bibinfo
  {author} {\bibfnamefont {F.}~\bibnamefont {{Galeazzi}}},\ }\bibfield  {title}
  {\bibinfo {title} {{How Loud Are Neutron Star Mergers?}},\ }\href
  {https://doi.org/10.1103/PhysRevD.94.024023} {\bibfield  {journal} {\bibinfo
  {journal} {Phys. Rev. D}\ }\textbf {\bibinfo {volume} {94}},\ \bibinfo {eid}
  {024023} (\bibinfo {year} {2016})},\ \Eprint
  {https://arxiv.org/abs/1512.06397} {arXiv:1512.06397 [gr-qc]} \BibitemShut
  {NoStop}%
\bibitem [{\citenamefont {{Shibata}}\ and\ \citenamefont
  {{Kiuchi}}(2017)}]{Shibata2017b}%
  \BibitemOpen
  \bibfield  {author} {\bibinfo {author} {\bibfnamefont {M.}~\bibnamefont
  {{Shibata}}}\ and\ \bibinfo {author} {\bibfnamefont {K.}~\bibnamefont
  {{Kiuchi}}},\ }\bibfield  {title} {\bibinfo {title} {{Gravitational waves
  from remnant massive neutron stars of binary neutron star merger: Viscous
  hydrodynamics effects}},\ }\href {https://doi.org/10.1103/PhysRevD.95.123003}
  {\bibfield  {journal} {\bibinfo  {journal} {Phys. Rev. D}\ }\textbf {\bibinfo
  {volume} {95}},\ \bibinfo {eid} {123003} (\bibinfo {year} {2017})},\ \Eprint
  {https://arxiv.org/abs/1705.06142} {arXiv:1705.06142 [astro-ph.HE]}
  \BibitemShut {NoStop}%
\bibitem [{\citenamefont {{Chaurasia}}\ \emph {et~al.}(2018)\citenamefont
  {{Chaurasia}}, \citenamefont {{Dietrich}}, \citenamefont
  {{Johnson-McDaniel}}, \citenamefont {{Ujevic}}, \citenamefont {{Tichy}},\
  and\ \citenamefont {{Br{\"u}gmann}}}]{Chaurasia2018}%
  \BibitemOpen
  \bibfield  {author} {\bibinfo {author} {\bibfnamefont {S.~V.}\ \bibnamefont
  {{Chaurasia}}}, \bibinfo {author} {\bibfnamefont {T.}~\bibnamefont
  {{Dietrich}}}, \bibinfo {author} {\bibfnamefont {N.~K.}\ \bibnamefont
  {{Johnson-McDaniel}}}, \bibinfo {author} {\bibfnamefont {M.}~\bibnamefont
  {{Ujevic}}}, \bibinfo {author} {\bibfnamefont {W.}~\bibnamefont {{Tichy}}},\
  and\ \bibinfo {author} {\bibfnamefont {B.}~\bibnamefont {{Br{\"u}gmann}}},\
  }\bibfield  {title} {\bibinfo {title} {{Gravitational waves and mass ejecta
  from binary neutron star mergers: Effect of large eccentricities}},\
  }\href@noop {} {\  (\bibinfo {year} {2018})},\ \Eprint
  {https://arxiv.org/abs/1807.06857} {arXiv:1807.06857 [gr-qc]} \BibitemShut
  {NoStop}%
\bibitem [{\citenamefont {{Stergioulas}}\ \emph {et~al.}(2011)\citenamefont
  {{Stergioulas}}, \citenamefont {{Bauswein}}, \citenamefont {{Zagkouris}},\
  and\ \citenamefont {{Janka}}}]{Stergioulas2011b}%
  \BibitemOpen
  \bibfield  {author} {\bibinfo {author} {\bibfnamefont {N.}~\bibnamefont
  {{Stergioulas}}}, \bibinfo {author} {\bibfnamefont {A.}~\bibnamefont
  {{Bauswein}}}, \bibinfo {author} {\bibfnamefont {K.}~\bibnamefont
  {{Zagkouris}}},\ and\ \bibinfo {author} {\bibfnamefont {H.-T.}\ \bibnamefont
  {{Janka}}},\ }\bibfield  {title} {\bibinfo {title} {{Gravitational waves and
  non-axisymmetric oscillation modes in mergers of compact object binaries}},\
  }\href {https://doi.org/10.1111/j.1365-2966.2011.19493.x} {\bibfield
  {journal} {\bibinfo  {journal} {Mon. Not. R. Astron. Soc.}\ }\textbf
  {\bibinfo {volume} {418}},\ \bibinfo {pages} {427} (\bibinfo {year}
  {2011})},\ \Eprint {https://arxiv.org/abs/1105.0368} {arXiv:1105.0368
  [gr-qc]} \BibitemShut {NoStop}%
\bibitem [{\citenamefont {{Bauswein}}\ and\ \citenamefont
  {{Janka}}(2012)}]{Bauswein2012a}%
  \BibitemOpen
  \bibfield  {author} {\bibinfo {author} {\bibfnamefont {A.}~\bibnamefont
  {{Bauswein}}}\ and\ \bibinfo {author} {\bibfnamefont {H.-T.}\ \bibnamefont
  {{Janka}}},\ }\bibfield  {title} {\bibinfo {title} {{Measuring Neutron-Star
  Properties via Gravitational Waves from Neutron-Star Mergers}},\ }\href
  {https://doi.org/10.1103/PhysRevLett.108.011101} {\bibfield  {journal}
  {\bibinfo  {journal} {Phys. Rev. Lett.}\ }\textbf {\bibinfo {volume} {108}},\
  \bibinfo {eid} {011101} (\bibinfo {year} {2012})},\ \Eprint
  {https://arxiv.org/abs/1106.1616} {arXiv:1106.1616 [astro-ph.SR]}
  \BibitemShut {NoStop}%
\bibitem [{\citenamefont {{Bauswein}}\ \emph {et~al.}(2012)\citenamefont
  {{Bauswein}}, \citenamefont {{Janka}}, \citenamefont {{Hebeler}},\ and\
  \citenamefont {{Schwenk}}}]{Bauswein2012}%
  \BibitemOpen
  \bibfield  {author} {\bibinfo {author} {\bibfnamefont {A.}~\bibnamefont
  {{Bauswein}}}, \bibinfo {author} {\bibfnamefont {H.-T.}\ \bibnamefont
  {{Janka}}}, \bibinfo {author} {\bibfnamefont {K.}~\bibnamefont {{Hebeler}}},\
  and\ \bibinfo {author} {\bibfnamefont {A.}~\bibnamefont {{Schwenk}}},\
  }\bibfield  {title} {\bibinfo {title} {{Equation-of-state dependence of the
  gravitational-wave signal from the ring-down phase of neutron-star
  mergers}},\ }\href {https://doi.org/10.1103/PhysRevD.86.063001} {\bibfield
  {journal} {\bibinfo  {journal} {Phys. Rev. D}\ }\textbf {\bibinfo {volume}
  {86}},\ \bibinfo {eid} {063001} (\bibinfo {year} {2012})},\ \Eprint
  {https://arxiv.org/abs/1204.1888} {arXiv:1204.1888 [astro-ph.SR]}
  \BibitemShut {NoStop}%
\bibitem [{\citenamefont {{Hotokezaka}}\ \emph {et~al.}(2013)\citenamefont
  {{Hotokezaka}}, \citenamefont {{Kiuchi}}, \citenamefont {{Kyutoku}},
  \citenamefont {{Muranushi}}, \citenamefont {{Sekiguchi}}, \citenamefont
  {{Shibata}},\ and\ \citenamefont {{Taniguchi}}}]{Hotokezaka2013c}%
  \BibitemOpen
  \bibfield  {author} {\bibinfo {author} {\bibfnamefont {K.}~\bibnamefont
  {{Hotokezaka}}}, \bibinfo {author} {\bibfnamefont {K.}~\bibnamefont
  {{Kiuchi}}}, \bibinfo {author} {\bibfnamefont {K.}~\bibnamefont {{Kyutoku}}},
  \bibinfo {author} {\bibfnamefont {T.}~\bibnamefont {{Muranushi}}}, \bibinfo
  {author} {\bibfnamefont {Y.-i.}\ \bibnamefont {{Sekiguchi}}}, \bibinfo
  {author} {\bibfnamefont {M.}~\bibnamefont {{Shibata}}},\ and\ \bibinfo
  {author} {\bibfnamefont {K.}~\bibnamefont {{Taniguchi}}},\ }\bibfield
  {title} {\bibinfo {title} {Remnant massive neutron stars of binary neutron
  star mergers: Evolution process and gravitational waveform},\ }\href
  {https://doi.org/10.1103/PhysRevD.88.044026} {\bibfield  {journal} {\bibinfo
  {journal} {Phys. Rev. D}\ }\textbf {\bibinfo {volume} {88}},\ \bibinfo {eid}
  {044026} (\bibinfo {year} {2013})},\ \Eprint
  {https://arxiv.org/abs/1307.5888} {arXiv:1307.5888 [astro-ph.HE]}
  \BibitemShut {NoStop}%
\bibitem [{\citenamefont {{Bauswein}}\ \emph {et~al.}(2014)\citenamefont
  {{Bauswein}}, \citenamefont {{Stergioulas}},\ and\ \citenamefont
  {{Janka}}}]{Bauswein2014}%
  \BibitemOpen
  \bibfield  {author} {\bibinfo {author} {\bibfnamefont {A.}~\bibnamefont
  {{Bauswein}}}, \bibinfo {author} {\bibfnamefont {N.}~\bibnamefont
  {{Stergioulas}}},\ and\ \bibinfo {author} {\bibfnamefont {H.-T.}\
  \bibnamefont {{Janka}}},\ }\bibfield  {title} {\bibinfo {title} {{Revealing
  the high-density equation of state through binary neutron star mergers}},\
  }\href {https://doi.org/10.1103/PhysRevD.90.023002} {\bibfield  {journal}
  {\bibinfo  {journal} {Phys. Rev. D}\ }\textbf {\bibinfo {volume} {90}},\
  \bibinfo {eid} {023002} (\bibinfo {year} {2014})},\ \Eprint
  {https://arxiv.org/abs/1403.5301} {arXiv:1403.5301 [astro-ph.SR]}
  \BibitemShut {NoStop}%
\bibitem [{\citenamefont {{Takami}}\ \emph {et~al.}(2015)\citenamefont
  {{Takami}}, \citenamefont {{Rezzolla}},\ and\ \citenamefont
  {{Baiotti}}}]{Takami2015}%
  \BibitemOpen
  \bibfield  {author} {\bibinfo {author} {\bibfnamefont {K.}~\bibnamefont
  {{Takami}}}, \bibinfo {author} {\bibfnamefont {L.}~\bibnamefont
  {{Rezzolla}}},\ and\ \bibinfo {author} {\bibfnamefont {L.}~\bibnamefont
  {{Baiotti}}},\ }\bibfield  {title} {\bibinfo {title} {{Spectral properties of
  the post-merger gravitational-wave signal from binary neutron stars}},\
  }\href {https://doi.org/10.1103/PhysRevD.91.064001} {\bibfield  {journal}
  {\bibinfo  {journal} {Phys. Rev. D}\ }\textbf {\bibinfo {volume} {91}},\
  \bibinfo {eid} {064001} (\bibinfo {year} {2015})},\ \Eprint
  {https://arxiv.org/abs/1412.3240} {arXiv:1412.3240 [gr-qc]} \BibitemShut
  {NoStop}%
\bibitem [{\citenamefont {{Rezzolla}}\ and\ \citenamefont
  {{Takami}}(2016)}]{Rezzolla2016}%
  \BibitemOpen
  \bibfield  {author} {\bibinfo {author} {\bibfnamefont {L.}~\bibnamefont
  {{Rezzolla}}}\ and\ \bibinfo {author} {\bibfnamefont {K.}~\bibnamefont
  {{Takami}}},\ }\bibfield  {title} {\bibinfo {title} {Gravitational-wave
  signal from binary neutron stars: A systematic analysis of the spectral
  properties},\ }\href {https://doi.org/10.1103/PhysRevD.93.124051} {\bibfield
  {journal} {\bibinfo  {journal} {Phys. Rev. D}\ }\textbf {\bibinfo {volume}
  {93}},\ \bibinfo {eid} {124051} (\bibinfo {year} {2016})},\ \Eprint
  {https://arxiv.org/abs/1604.00246} {arXiv:1604.00246 [gr-qc]} \BibitemShut
  {NoStop}%
\bibitem [{\citenamefont {Bernuzzi}\ \emph {et~al.}(2015)\citenamefont
  {Bernuzzi}, \citenamefont {Dietrich},\ and\ \citenamefont
  {Nagar}}]{Bernuzzi:2015rla}%
  \BibitemOpen
  \bibfield  {author} {\bibinfo {author} {\bibfnamefont {S.}~\bibnamefont
  {Bernuzzi}}, \bibinfo {author} {\bibfnamefont {T.}~\bibnamefont {Dietrich}},\
  and\ \bibinfo {author} {\bibfnamefont {A.}~\bibnamefont {Nagar}},\ }\bibfield
   {title} {\bibinfo {title} {{Modeling the complete gravitational wave
  spectrum of neutron star mergers}},\ }\href
  {https://doi.org/10.1103/PhysRevLett.115.091101} {\bibfield  {journal}
  {\bibinfo  {journal} {Phys. Rev. Lett.}\ }\textbf {\bibinfo {volume} {115}},\
  \bibinfo {pages} {091101} (\bibinfo {year} {2015})},\ \Eprint
  {https://arxiv.org/abs/1504.01764} {arXiv:1504.01764 [gr-qc]} \BibitemShut
  {NoStop}%
\bibitem [{\citenamefont {Bauswein}\ and\ \citenamefont
  {Stergioulas}(2015)}]{Bauswein:2015yca}%
  \BibitemOpen
  \bibfield  {author} {\bibinfo {author} {\bibfnamefont {A.}~\bibnamefont
  {Bauswein}}\ and\ \bibinfo {author} {\bibfnamefont {N.}~\bibnamefont
  {Stergioulas}},\ }\bibfield  {title} {\bibinfo {title} {{Unified picture of
  the post-merger dynamics and gravitational wave emission in neutron star
  mergers}},\ }\href {https://doi.org/10.1103/PhysRevD.91.124056} {\bibfield
  {journal} {\bibinfo  {journal} {Phys. Rev.}\ }\textbf {\bibinfo {volume}
  {D91}},\ \bibinfo {pages} {124056} (\bibinfo {year} {2015})},\ \Eprint
  {https://arxiv.org/abs/1502.03176} {arXiv:1502.03176 [astro-ph.SR]}
  \BibitemShut {NoStop}%
\bibitem [{\citenamefont {{Nakamura}}\ \emph {et~al.}(1987)\citenamefont
  {{Nakamura}}, \citenamefont {{Oohara}},\ and\ \citenamefont
  {{Kojima}}}]{Nakamura87}%
  \BibitemOpen
  \bibfield  {author} {\bibinfo {author} {\bibfnamefont {T.}~\bibnamefont
  {{Nakamura}}}, \bibinfo {author} {\bibfnamefont {K.}~\bibnamefont
  {{Oohara}}},\ and\ \bibinfo {author} {\bibfnamefont {Y.}~\bibnamefont
  {{Kojima}}},\ }\bibfield  {title} {\bibinfo {title} {{General Relativistic
  Collapse to Black Holes and Gravitational Waves from Black Holes}},\ }\href
  {https://doi.org/10.1143/PTPS.90.1} {\bibfield  {journal} {\bibinfo
  {journal} {Progress of Theoretical Physics Supplement}\ }\textbf {\bibinfo
  {volume} {90}},\ \bibinfo {pages} {1} (\bibinfo {year} {1987})}\BibitemShut
  {NoStop}%
\bibitem [{\citenamefont {{Shibata}}\ and\ \citenamefont
  {{Nakamura}}(1995)}]{Shibata95}%
  \BibitemOpen
  \bibfield  {author} {\bibinfo {author} {\bibfnamefont {M.}~\bibnamefont
  {{Shibata}}}\ and\ \bibinfo {author} {\bibfnamefont {T.}~\bibnamefont
  {{Nakamura}}},\ }\bibfield  {title} {\bibinfo {title} {{Evolution of
  three-dimensional gravitational waves: Harmonic slicing case}},\ }\href
  {https://doi.org/10.1103/PhysRevD.52.5428} {\bibfield  {journal} {\bibinfo
  {journal} {Phys. Rev. D}\ }\textbf {\bibinfo {volume} {52}},\ \bibinfo
  {pages} {5428} (\bibinfo {year} {1995})}\BibitemShut {NoStop}%
\bibitem [{\citenamefont {{Baumgarte}}\ and\ \citenamefont
  {{Shapiro}}(1999)}]{Baumgarte99}%
  \BibitemOpen
  \bibfield  {author} {\bibinfo {author} {\bibfnamefont {T.~W.}\ \bibnamefont
  {{Baumgarte}}}\ and\ \bibinfo {author} {\bibfnamefont {S.~L.}\ \bibnamefont
  {{Shapiro}}},\ }\bibfield  {title} {\bibinfo {title} {{Numerical integration
  of Einstein's field equations}},\ }\href
  {https://doi.org/10.1103/PhysRevD.59.024007} {\bibfield  {journal} {\bibinfo
  {journal} {Phys. Rev. D}\ }\textbf {\bibinfo {volume} {59}},\ \bibinfo {eid}
  {024007} (\bibinfo {year} {1999})},\ \Eprint
  {https://arxiv.org/abs/gr-qc/9810065} {gr-qc/9810065} \BibitemShut {NoStop}%
\bibitem [{\citenamefont {Brown}(2009)}]{Brown09}%
  \BibitemOpen
  \bibfield  {author} {\bibinfo {author} {\bibfnamefont {D.~J.}\ \bibnamefont
  {Brown}},\ }\bibfield  {title} {\bibinfo {title} {{Covariant formulations of
  Baumgarte, Shapiro, Shibata, and Nakamura and the standard gauge}},\ }\href
  {https://doi.org/10.1103/PhysRevD.79.104029} {\bibfield  {journal} {\bibinfo
  {journal} {Phys. Rev. D}\ }\textbf {\bibinfo {volume} {79}},\ \bibinfo
  {pages} {104029} (\bibinfo {year} {2009})}\BibitemShut {NoStop}%
\bibitem [{\citenamefont {Harten}\ \emph {et~al.}(1983)\citenamefont {Harten},
  \citenamefont {Lax},\ and\ \citenamefont {van Leer}}]{Harten83}%
  \BibitemOpen
  \bibfield  {author} {\bibinfo {author} {\bibfnamefont {A.}~\bibnamefont
  {Harten}}, \bibinfo {author} {\bibfnamefont {P.~D.}\ \bibnamefont {Lax}},\
  and\ \bibinfo {author} {\bibfnamefont {B.}~\bibnamefont {van Leer}},\
  }\bibfield  {title} {\bibinfo {title} {On upstream differencing and
  godunov-type schemes for hyperbolic conservation laws},\ }\href
  {https://doi.org/10.1137/1025002} {\bibfield  {journal} {\bibinfo  {journal}
  {SIAM Rev.}\ }\textbf {\bibinfo {volume} {25}},\ \bibinfo {pages} {35}
  (\bibinfo {year} {1983})}\BibitemShut {NoStop}%
\bibitem [{\citenamefont {Colella}\ and\ \citenamefont
  {Woodward}(1984)}]{Colella84}%
  \BibitemOpen
  \bibfield  {author} {\bibinfo {author} {\bibfnamefont {P.}~\bibnamefont
  {Colella}}\ and\ \bibinfo {author} {\bibfnamefont {P.~R.}\ \bibnamefont
  {Woodward}},\ }\bibfield  {title} {\bibinfo {title} {The piecewise parabolic
  method (ppm) for gas-dynamical simulations},\ }\href {https://doi.org/DOI:
  10.1016/0021-9991(84)90143-8} {\bibfield  {journal} {\bibinfo  {journal}
  {Journal of Computational Physics}\ }\textbf {\bibinfo {volume} {54}},\
  \bibinfo {pages} {174} (\bibinfo {year} {1984})}\BibitemShut {NoStop}%
\bibitem [{\citenamefont {{Alford}}\ \emph {et~al.}(2005)\citenamefont
  {{Alford}}, \citenamefont {{Braby}}, \citenamefont {{Paris}},\ and\
  \citenamefont {{Reddy}}}]{Alford2005}%
  \BibitemOpen
  \bibfield  {author} {\bibinfo {author} {\bibfnamefont {M.}~\bibnamefont
  {{Alford}}}, \bibinfo {author} {\bibfnamefont {M.}~\bibnamefont {{Braby}}},
  \bibinfo {author} {\bibfnamefont {M.}~\bibnamefont {{Paris}}},\ and\ \bibinfo
  {author} {\bibfnamefont {S.}~\bibnamefont {{Reddy}}},\ }\bibfield  {title}
  {\bibinfo {title} {Hybrid stars that masquerade as neutron stars},\ }\href
  {https://doi.org/10.1086/430902} {\bibfield  {journal} {\bibinfo  {journal}
  {Astrophys. J.}\ }\textbf {\bibinfo {volume} {629}},\ \bibinfo {pages} {969}
  (\bibinfo {year} {2005})},\ \Eprint {https://arxiv.org/abs/nucl-th/0411016}
  {nucl-th/0411016} \BibitemShut {NoStop}%
\bibitem [{\citenamefont {{Akmal}}\ \emph {et~al.}(1998)\citenamefont
  {{Akmal}}, \citenamefont {{Pandharipande}},\ and\ \citenamefont
  {{Ravenhall}}}]{Akmal1998}%
  \BibitemOpen
  \bibfield  {author} {\bibinfo {author} {\bibfnamefont {A.}~\bibnamefont
  {{Akmal}}}, \bibinfo {author} {\bibfnamefont {V.~R.}\ \bibnamefont
  {{Pandharipande}}},\ and\ \bibinfo {author} {\bibfnamefont {D.~G.}\
  \bibnamefont {{Ravenhall}}},\ }\bibfield  {title} {\bibinfo {title}
  {{Equation of state of nucleon matter and neutron star structure}},\ }\href
  {https://doi.org/10.1103/PhysRevC.58.1804} {\bibfield  {journal} {\bibinfo
  {journal} {Phys. Rev. C}\ }\textbf {\bibinfo {volume} {58}},\ \bibinfo
  {pages} {1804} (\bibinfo {year} {1998})},\ \Eprint
  {https://arxiv.org/abs/arXiv:hep-ph/9804388} {arXiv:hep-ph/9804388}
  \BibitemShut {NoStop}%
\bibitem [{\citenamefont {{Glendenning}}(1985)}]{Glendenning1985}%
  \BibitemOpen
  \bibfield  {author} {\bibinfo {author} {\bibfnamefont {N.~K.}\ \bibnamefont
  {{Glendenning}}},\ }\bibfield  {title} {\bibinfo {title} {Neutron stars are
  giant hypernuclei?},\ }\href {https://doi.org/10.1086/163253} {\bibfield
  {journal} {\bibinfo  {journal} {Astrophys. J.}\ }\textbf {\bibinfo {volume}
  {293}},\ \bibinfo {pages} {470} (\bibinfo {year} {1985})}\BibitemShut
  {NoStop}%
\bibitem [{\citenamefont {Glendenning}\ \emph {et~al.}(1992)\citenamefont
  {Glendenning}, \citenamefont {Weber},\ and\ \citenamefont
  {Moszkowski}}]{Glendenning1992}%
  \BibitemOpen
  \bibfield  {author} {\bibinfo {author} {\bibfnamefont {N.}~\bibnamefont
  {Glendenning}}, \bibinfo {author} {\bibfnamefont {F.}~\bibnamefont {Weber}},\
  and\ \bibinfo {author} {\bibfnamefont {S.}~\bibnamefont {Moszkowski}},\
  }\bibfield  {title} {\bibinfo {title} {Neutron stars in the derivative
  coupling model},\ }\href@noop {} {\bibfield  {journal} {\bibinfo  {journal}
  {Physical Review C}\ }\textbf {\bibinfo {volume} {45}},\ \bibinfo {pages}
  {844} (\bibinfo {year} {1992})}\BibitemShut {NoStop}%
\bibitem [{\citenamefont {{Douchin}}\ and\ \citenamefont
  {{Haensel}}(2001)}]{Douchin2001}%
  \BibitemOpen
  \bibfield  {author} {\bibinfo {author} {\bibfnamefont {F.}~\bibnamefont
  {{Douchin}}}\ and\ \bibinfo {author} {\bibfnamefont {P.}~\bibnamefont
  {{Haensel}}},\ }\bibfield  {title} {\bibinfo {title} {{A unified equation of
  state of dense matter and neutron star structure}},\ }\href
  {https://doi.org/10.1051/0004-6361:20011402} {\bibfield  {journal} {\bibinfo
  {journal} {Astron. Astrophys.}\ }\textbf {\bibinfo {volume} {380}},\ \bibinfo
  {pages} {151} (\bibinfo {year} {2001})},\ \Eprint
  {https://arxiv.org/abs/arXiv:astro-ph/0111092} {arXiv:astro-ph/0111092}
  \BibitemShut {NoStop}%
\bibitem [{\citenamefont {Antoniadis}\ \emph {et~al.}(2013)\citenamefont
  {Antoniadis} \emph {et~al.}}]{Antoniadis1233232}%
  \BibitemOpen
  \bibfield  {author} {\bibinfo {author} {\bibfnamefont {J.}~\bibnamefont
  {Antoniadis}} \emph {et~al.},\ }\bibfield  {title} {\bibinfo {title} {A
  massive pulsar in a compact relativistic binary},\ }\bibfield  {journal}
  {\bibinfo  {journal} {Science}\ }\textbf {\bibinfo {volume} {340}},\ \href
  {https://doi.org/10.1126/science.1233232} {10.1126/science.1233232} (\bibinfo
  {year} {2013})\BibitemShut {NoStop}%
\bibitem [{\citenamefont {Cromartie}\ \emph {et~al.}(2019)\citenamefont
  {Cromartie} \emph {et~al.}}]{Cromartie:2019kug}%
  \BibitemOpen
  \bibfield  {author} {\bibinfo {author} {\bibfnamefont {H.~T.}\ \bibnamefont
  {Cromartie}} \emph {et~al.},\ }\bibfield  {title} {\bibinfo {title}
  {{Relativistic Shapiro delay measurements of an extremely massive millisecond
  pulsar}},\ }\href {https://doi.org/10.1038/s41550-019-0880-2} {\bibfield
  {journal} {\bibinfo  {journal} {Nature Astron.}\ }\textbf {\bibinfo {volume}
  {4}},\ \bibinfo {pages} {72} (\bibinfo {year} {2019})},\ \Eprint
  {https://arxiv.org/abs/1904.06759} {arXiv:1904.06759 [astro-ph.HE]}
  \BibitemShut {NoStop}%
\bibitem [{\citenamefont {Linares}\ \emph {et~al.}(2018)\citenamefont
  {Linares}, \citenamefont {Shahbaz},\ and\ \citenamefont
  {Casares}}]{Linares_2018}%
  \BibitemOpen
  \bibfield  {author} {\bibinfo {author} {\bibfnamefont {M.}~\bibnamefont
  {Linares}}, \bibinfo {author} {\bibfnamefont {T.}~\bibnamefont {Shahbaz}},\
  and\ \bibinfo {author} {\bibfnamefont {J.}~\bibnamefont {Casares}},\
  }\bibfield  {title} {\bibinfo {title} {Peering into the dark side: Magnesium
  lines establish a massive neutron star in {PSR} j2215+5135},\ }\href
  {https://doi.org/10.3847/1538-4357/aabde6} {\bibfield  {journal} {\bibinfo
  {journal} {The Astrophysical Journal}\ }\textbf {\bibinfo {volume} {859}},\
  \bibinfo {pages} {54} (\bibinfo {year} {2018})}\BibitemShut {NoStop}%
\bibitem [{\citenamefont {Godzieba}\ \emph {et~al.}(2020)\citenamefont
  {Godzieba}, \citenamefont {Radice},\ and\ \citenamefont
  {Bernuzzi}}]{Godzieba:2020tjn}%
  \BibitemOpen
  \bibfield  {author} {\bibinfo {author} {\bibfnamefont {D.~A.}\ \bibnamefont
  {Godzieba}}, \bibinfo {author} {\bibfnamefont {D.}~\bibnamefont {Radice}},\
  and\ \bibinfo {author} {\bibfnamefont {S.}~\bibnamefont {Bernuzzi}},\
  }\bibfield  {title} {\bibinfo {title} {{On the maximum mass of neutron stars
  and GW190814}},\ }\href@noop {} {\  (\bibinfo {year} {2020})},\ \Eprint
  {https://arxiv.org/abs/2007.10999} {arXiv:2007.10999 [astro-ph.HE]}
  \BibitemShut {NoStop}%
\bibitem [{\citenamefont {{Baumgarte}}\ and\ \citenamefont
  {{Shapiro}}(2010)}]{Baumgarte2010a}%
  \BibitemOpen
  \bibfield  {author} {\bibinfo {author} {\bibfnamefont {T.~W.}\ \bibnamefont
  {{Baumgarte}}}\ and\ \bibinfo {author} {\bibfnamefont {S.~L.}\ \bibnamefont
  {{Shapiro}}},\ }\href@noop {} {\emph {\bibinfo {title} {{Numerical
  Relativity: Solving Einstein's Equations on the Computer}}}}\ (\bibinfo
  {publisher} {Cambridge University Press},\ \bibinfo {address} {Cambridge,
  UK},\ \bibinfo {year} {2010})\BibitemShut {NoStop}%
\bibitem [{\citenamefont {Maggiore}(2007)}]{Maggiore2007}%
  \BibitemOpen
  \bibfield  {author} {\bibinfo {author} {\bibfnamefont {M.}~\bibnamefont
  {Maggiore}},\ }\href {http://books.google.de/books?id=AqVpQgAACAAJ} {\emph
  {\bibinfo {title} {Gravitational Waves: Volume 1: Theory and Experiments}}},\
  Gravitational Waves\ (\bibinfo  {publisher} {Oxford University Press, USA},\
  \bibinfo {year} {2007})\BibitemShut {NoStop}%
\bibitem [{\citenamefont {Agathos}\ \emph {et~al.}(2020)\citenamefont
  {Agathos}, \citenamefont {Zappa}, \citenamefont {Bernuzzi}, \citenamefont
  {Perego}, \citenamefont {Breschi},\ and\ \citenamefont
  {Radice}}]{Agathos:2019sah}%
  \BibitemOpen
  \bibfield  {author} {\bibinfo {author} {\bibfnamefont {M.}~\bibnamefont
  {Agathos}}, \bibinfo {author} {\bibfnamefont {F.}~\bibnamefont {Zappa}},
  \bibinfo {author} {\bibfnamefont {S.}~\bibnamefont {Bernuzzi}}, \bibinfo
  {author} {\bibfnamefont {A.}~\bibnamefont {Perego}}, \bibinfo {author}
  {\bibfnamefont {M.}~\bibnamefont {Breschi}},\ and\ \bibinfo {author}
  {\bibfnamefont {D.}~\bibnamefont {Radice}},\ }\bibfield  {title} {\bibinfo
  {title} {{Inferring Prompt Black-Hole Formation in Neutron Star Mergers from
  Gravitational-Wave Data}},\ }\href
  {https://doi.org/10.1103/PhysRevD.101.044006} {\bibfield  {journal} {\bibinfo
   {journal} {Phys. Rev. D}\ }\textbf {\bibinfo {volume} {101}},\ \bibinfo
  {pages} {044006} (\bibinfo {year} {2020})},\ \Eprint
  {https://arxiv.org/abs/1908.05442} {arXiv:1908.05442 [gr-qc]} \BibitemShut
  {NoStop}%
\bibitem [{\citenamefont {Kiziltan}\ \emph {et~al.}(2013)\citenamefont
  {Kiziltan}, \citenamefont {Kottas}, \citenamefont {De~Yoreo},\ and\
  \citenamefont {Thorsett}}]{Kiziltan2013}%
  \BibitemOpen
  \bibfield  {author} {\bibinfo {author} {\bibfnamefont {B.}~\bibnamefont
  {Kiziltan}}, \bibinfo {author} {\bibfnamefont {A.}~\bibnamefont {Kottas}},
  \bibinfo {author} {\bibfnamefont {M.}~\bibnamefont {De~Yoreo}},\ and\
  \bibinfo {author} {\bibfnamefont {S.~E.}\ \bibnamefont {Thorsett}},\
  }\bibfield  {title} {\bibinfo {title} {The neutron star mass distribution},\
  }\href {https://doi.org/10.1088/0004-637X/778/1/66} {\bibfield  {journal}
  {\bibinfo  {journal} {Astrophys. J.}\ }\textbf {\bibinfo {volume} {778}},\
  \bibinfo {pages} {66} (\bibinfo {year} {2013})},\ \Eprint
  {https://arxiv.org/abs/1309.6635} {arXiv:1309.6635 [astro-ph.SR]}
  \BibitemShut {NoStop}%
\bibitem [{\citenamefont {{Bovard}}\ \emph {et~al.}(2017)\citenamefont
  {{Bovard}}, \citenamefont {{Martin}}, \citenamefont {{Guercilena}},
  \citenamefont {{Arcones}}, \citenamefont {{Rezzolla}},\ and\ \citenamefont
  {{Korobkin}}}]{Bovard2017}%
  \BibitemOpen
  \bibfield  {author} {\bibinfo {author} {\bibfnamefont {L.}~\bibnamefont
  {{Bovard}}}, \bibinfo {author} {\bibfnamefont {D.}~\bibnamefont {{Martin}}},
  \bibinfo {author} {\bibfnamefont {F.}~\bibnamefont {{Guercilena}}}, \bibinfo
  {author} {\bibfnamefont {A.}~\bibnamefont {{Arcones}}}, \bibinfo {author}
  {\bibfnamefont {L.}~\bibnamefont {{Rezzolla}}},\ and\ \bibinfo {author}
  {\bibfnamefont {O.}~\bibnamefont {{Korobkin}}},\ }\bibfield  {title}
  {\bibinfo {title} {{On r-process nucleosynthesis from matter ejected in
  binary neutron star mergers}},\ }\href@noop {} {\bibfield  {journal}
  {\bibinfo  {journal} {Phys. Rev. D}\ }\textbf {\bibinfo {volume} {96}},\
  \bibinfo {pages} {124005} (\bibinfo {year} {2017})},\ \Eprint
  {https://arxiv.org/abs/1709.09630} {arXiv:1709.09630 [gr-qc]} \BibitemShut
  {NoStop}%
\bibitem [{\citenamefont {Lucca}\ and\ \citenamefont
  {Sagunski}(2020)}]{Lucca:2019ohp}%
  \BibitemOpen
  \bibfield  {author} {\bibinfo {author} {\bibfnamefont {M.}~\bibnamefont
  {Lucca}}\ and\ \bibinfo {author} {\bibfnamefont {L.}~\bibnamefont
  {Sagunski}},\ }\bibfield  {title} {\bibinfo {title} {{The lifetime of binary
  neutron star merger remnants}},\ }\href
  {https://doi.org/10.1016/j.jheap.2020.04.003} {\bibfield  {journal} {\bibinfo
   {journal} {JHEAp}\ }\textbf {\bibinfo {volume} {27}},\ \bibinfo {pages} {33}
  (\bibinfo {year} {2020})},\ \Eprint {https://arxiv.org/abs/1909.08631}
  {arXiv:1909.08631 [astro-ph.HE]} \BibitemShut {NoStop}%
\bibitem [{\citenamefont {Bauswein}\ \emph {et~al.}(2010)\citenamefont
  {Bauswein}, \citenamefont {Janka},\ and\ \citenamefont
  {Oechslin}}]{Bauswein2010Testing}%
  \BibitemOpen
  \bibfield  {author} {\bibinfo {author} {\bibfnamefont {A.}~\bibnamefont
  {Bauswein}}, \bibinfo {author} {\bibfnamefont {H.-T.}\ \bibnamefont
  {Janka}},\ and\ \bibinfo {author} {\bibfnamefont {R.}~\bibnamefont
  {Oechslin}},\ }\bibfield  {title} {\bibinfo {title} {Testing approximations
  of thermal effects in neutron star merger simulations},\ }\href
  {https://doi.org/10.1103/PhysRevD.82.084043} {\bibfield  {journal} {\bibinfo
  {journal} {Phys. Rev. D}\ }\textbf {\bibinfo {volume} {82}},\ \bibinfo
  {pages} {084043} (\bibinfo {year} {2010})}\BibitemShut {NoStop}%
\bibitem [{\citenamefont {Shibata}\ and\ \citenamefont
  {Hotokezaka}(2019)}]{Shibata2019Merger}%
  \BibitemOpen
  \bibfield  {author} {\bibinfo {author} {\bibfnamefont {M.}~\bibnamefont
  {Shibata}}\ and\ \bibinfo {author} {\bibfnamefont {K.}~\bibnamefont
  {Hotokezaka}},\ }\bibfield  {title} {\bibinfo {title} {Merger and mass
  ejection of neutron star binaries},\ }\href
  {https://doi.org/10.1146/annurev-nucl-101918-023625} {\bibfield  {journal}
  {\bibinfo  {journal} {Annual Review of Nuclear and Particle Science}\
  }\textbf {\bibinfo {volume} {69}},\ \bibinfo {pages} {41} (\bibinfo {year}
  {2019})}\BibitemShut {NoStop}%
\bibitem [{\citenamefont {Ciolfi}\ \emph {et~al.}(2019)\citenamefont {Ciolfi},
  \citenamefont {Kastaun}, \citenamefont {Kalinani},\ and\ \citenamefont
  {Giacomazzo}}]{Ciolfi:2019fie}%
  \BibitemOpen
  \bibfield  {author} {\bibinfo {author} {\bibfnamefont {R.}~\bibnamefont
  {Ciolfi}}, \bibinfo {author} {\bibfnamefont {W.}~\bibnamefont {Kastaun}},
  \bibinfo {author} {\bibfnamefont {J.~V.}\ \bibnamefont {Kalinani}},\ and\
  \bibinfo {author} {\bibfnamefont {B.}~\bibnamefont {Giacomazzo}},\ }\bibfield
   {title} {\bibinfo {title} {{First 100 ms of a long-lived magnetized neutron
  star formed in a binary neutron star merger}},\ }\href
  {https://doi.org/10.1103/PhysRevD.100.023005} {\bibfield  {journal} {\bibinfo
   {journal} {Phys. Rev. D}\ }\textbf {\bibinfo {volume} {100}},\ \bibinfo
  {pages} {023005} (\bibinfo {year} {2019})},\ \Eprint
  {https://arxiv.org/abs/1904.10222} {arXiv:1904.10222 [astro-ph.HE]}
  \BibitemShut {NoStop}%
\bibitem [{\citenamefont {Ciolfi}(2020)}]{Ciolfi:2020hgg}%
  \BibitemOpen
  \bibfield  {author} {\bibinfo {author} {\bibfnamefont {R.}~\bibnamefont
  {Ciolfi}},\ }\bibfield  {title} {\bibinfo {title} {{Collimated outflows from
  long-lived binary neutron star merger remnants}},\ }\href
  {https://doi.org/10.1093/mnrasl/slaa062} {\bibfield  {journal} {\bibinfo
  {journal} {Mon. Not. Roy. Astron. Soc.}\ }\textbf {\bibinfo {volume} {495}},\
  \bibinfo {pages} {L66} (\bibinfo {year} {2020})},\ \Eprint
  {https://arxiv.org/abs/2001.10241} {arXiv:2001.10241 [astro-ph.HE]}
  \BibitemShut {NoStop}%
\bibitem [{\citenamefont {Breschi}\ \emph {et~al.}(2019)\citenamefont
  {Breschi}, \citenamefont {Bernuzzi}, \citenamefont {Zappa}, \citenamefont
  {Agathos}, \citenamefont {Perego}, \citenamefont {Radice},\ and\
  \citenamefont {Nagar}}]{Breschi20}%
  \BibitemOpen
  \bibfield  {author} {\bibinfo {author} {\bibfnamefont {M.}~\bibnamefont
  {Breschi}}, \bibinfo {author} {\bibfnamefont {S.}~\bibnamefont {Bernuzzi}},
  \bibinfo {author} {\bibfnamefont {F.}~\bibnamefont {Zappa}}, \bibinfo
  {author} {\bibfnamefont {M.}~\bibnamefont {Agathos}}, \bibinfo {author}
  {\bibfnamefont {A.}~\bibnamefont {Perego}}, \bibinfo {author} {\bibfnamefont
  {D.}~\bibnamefont {Radice}},\ and\ \bibinfo {author} {\bibfnamefont
  {A.}~\bibnamefont {Nagar}},\ }\bibfield  {title} {\bibinfo {title} {Kilohertz
  gravitational waves from binary neutron star remnants: Time-domain model and
  constraints on extreme matter},\ }\href
  {https://doi.org/10.1103/PhysRevD.100.104029} {\bibfield  {journal} {\bibinfo
   {journal} {Phys. Rev. D}\ }\textbf {\bibinfo {volume} {100}},\ \bibinfo
  {pages} {104029} (\bibinfo {year} {2019})}\BibitemShut {NoStop}%
\bibitem [{\citenamefont {Perez}\ \emph {et~al.}(2012)\citenamefont {Perez},
  \citenamefont {Jansen},\ and\ \citenamefont {Martins}}]{perez2012}%
  \BibitemOpen
  \bibfield  {author} {\bibinfo {author} {\bibfnamefont {R.~E.}\ \bibnamefont
  {Perez}}, \bibinfo {author} {\bibfnamefont {P.~W.}\ \bibnamefont {Jansen}},\
  and\ \bibinfo {author} {\bibfnamefont {J.~R.}\ \bibnamefont {Martins}},\
  }\bibfield  {title} {\bibinfo {title} {pyopt: a python-based object-oriented
  framework for nonlinear constrained optimization},\ }\href@noop {} {\bibfield
   {journal} {\bibinfo  {journal} {Structural and Multidisciplinary
  Optimization}\ }\textbf {\bibinfo {volume} {45}},\ \bibinfo {pages} {101}
  (\bibinfo {year} {2012})}\BibitemShut {NoStop}%
\bibitem [{\citenamefont {Deb}\ \emph {et~al.}(2002)\citenamefont {Deb},
  \citenamefont {Pratap}, \citenamefont {Agarwal},\ and\ \citenamefont
  {Meyarivan}}]{deb2002}%
  \BibitemOpen
  \bibfield  {author} {\bibinfo {author} {\bibfnamefont {K.}~\bibnamefont
  {Deb}}, \bibinfo {author} {\bibfnamefont {A.}~\bibnamefont {Pratap}},
  \bibinfo {author} {\bibfnamefont {S.}~\bibnamefont {Agarwal}},\ and\ \bibinfo
  {author} {\bibfnamefont {T.}~\bibnamefont {Meyarivan}},\ }\bibfield  {title}
  {\bibinfo {title} {{A fast and elitist multiobjective genetic algorithm:
  NSGA-II}},\ }\href@noop {} {\bibfield  {journal} {\bibinfo  {journal} {IEEE
  transactions on evolutionary computation}\ }\textbf {\bibinfo {volume} {6}},\
  \bibinfo {pages} {182} (\bibinfo {year} {2002})}\BibitemShut {NoStop}%
\end{thebibliography}%

\end{document}